# digHolo

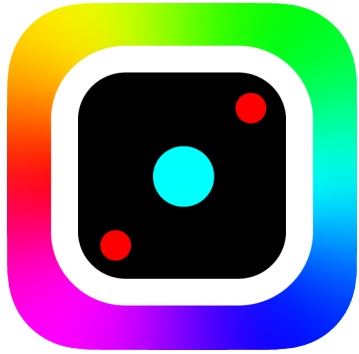

# User Guide

Version 1.0

April 1 2022

**Video Introduction** https://youtu.be/QLAqHMhsWXk

**Code** https://github.com/joelacarpenter/digHolo

i

# Background

## off-axis digital holography

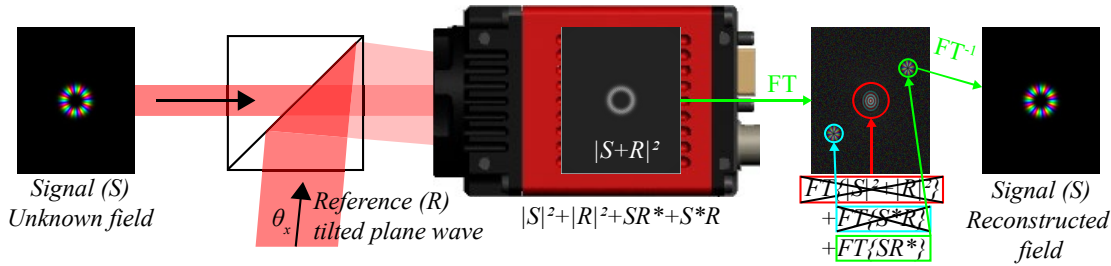

*Signal (S) Unknown field*    $\theta_x$ *Reference (R) tilted plane wave*    $|S+R|^2$    $|S|^2+|R|^2+SR^*+S^*R$    $FT\{|S|^2+|R|^2\}$ $+FT\{S^*R\}$ $+FT\{SR^*\}$    *Signal (S) Reconstructed field*

Off-axis digital holography is a technique for measuring the spatial amplitude and phase of an unknown field (S) by interfering the field with a known plane-wave reference field (R) travelling at an angle offset (off-axis) relative to the unknown field (S). The intensity of the interference pattern between the two beams, S and R, is recorded on a camera and then processed to recover the unknown field S. The process is similar to traditional off-axis holography, except the interference pattern is recorded on a digital camera (rather than film), and the operations of diffraction and spatial filtering are performed numerically rather than physically. It is a relatively straightforward procedure, whereby an interference pattern is Fourier transformed, a single term/area is selected in Fourier space, rejecting all else, and then inverse Fourier transformed to recover the desired Signal field, S. The figure below illustrates the procedure by which the interference pattern is measured and processed. Our unknown Signal field (S) and our off-axis Reference field (R) interfere on a camera, which records the intensity of the interference pattern of the Signal and Reference, $|S+R|^2$. The Signal and Reference fields must be mutually coherent on the timescale of the exposure time of the camera. That is, the relative phase between the Signal and Reference must be fixed and stable while the camera frame is being captured, otherwise the fringes on the interference pattern $|S+R|^2$ will blur out and it will not be possible to reconstruct the Signal. That is, we can not reconstruct the phase of the Signal relative to the Reference, if indeed there is no phase relationship between the Signal and Reference as far as the captured camera frame is concerned. The interference pattern $|S+R|^2$ can be expanded and rewritten as the summation of four terms, $|S|^2+|R|^2+SR^*+R^*S$. If we take the Fourier transform (FT) of the interference pattern, we end up with a summation of the Fourier transform of each of those four terms $FT\{|S|^2\}+FT\{|R|^2\}+FT\{SR^*\}+FT\{R^*S\}$, which in the figure is illustrated to the right of the camera. The Fourier transform of the intensity of the Reference $FT\{|R|^2\}$ and the Fourier transform of the intensity of the Signal, $FT\{|S|^2\}$, are autocorrelation terms that show up on-axis in Fourier-space indicated in the red circle. These do not contain any information about the relative phase between the Signal and Reference and are not of use. The off-axis terms indicated in green and blue are where the useful information lies. Indicated in green is the term $FT\{SR^*\}$, which is the convolution of the Fourier transform of S and the Fourier transform of R. As the Reference is a plane-wave, it's Fourier transform is a single point in Fourier-space at the position corresponding to theta_x. A convolution of a function with a delta, is simply the function offset to the position of the delta (sifting property). That is, the term $FT\{SR^*\}$ is just $FT\{S\}$ where S has been offset in Fourier space by the angle of the Reference field. Hence if we select this term, and then inverse Fourier transform back into the plane of the camera, and then remove the angle offset of the Reference wave, we get the reconstructed Signal field. The angle offset of the Reference wave can be removed either by shifting the $FT\{SR^*\}$ term to the centre of Fourier space before applying the inverse Fourier transform, or equivalently, by multiplying the reconstructed field by $R^*$ after the inverse Fourier transform.

*How do you choose the angle of the Reference Field?*

The incident angle must be sufficiently large that the autocorrelation terms and the cross-correlation terms can be separated in Fourier space, but not so large that cross-correlation terms do not fit in the resolvable Fourier space. That is, there must be no features of the interference pattern that change on a scale smaller than a single pixel of the camera.

From the figure, the auto-correlation terms illustrated in the red window and the cross-correlation terms of the green/blue regions must not overlap. The auto-correlation term will have twice the width of the cross-correlation term in Fourier space, and hence the Reference angle must be at least three times the maximum angle (spatial frequency) present in the Signal field. If the spatial frequency content of the Signal field has a radius of $w_c$ in Fourier space (green window), then the maximum possible Reference field angle would be given by $\sim 1/3$ $[sqrt(2)/(3+sqrt(2))]=0.320513$ of the maximum resolvable angle set by the frame pixel size ($w_{max}$). For example, if the wavelength is 1565e-9 and the pixelSize is 20e-6, $w_{max}$ would be $(1565e-9/(2*20e-6))*(180/pi) = 2.24$ degrees, and window radius ($w_c$) should be less than $0.3205*2.24=0.719$ degrees. The reference beam tilt angle in x and y would have to be $(x,y)=(3w_c,3w_c)/sqrt(2)=(1.525,1.525)$ degrees. If the full resolution of the camera is not required, smaller windows and shallower reference angles can be employed. Smaller windows are also less likely to capture unwanted noise. If the green window is allowed to wrap-around along axis in Fourier-space, a larger fractional window could be used [+ root of $0=8w^2+2-2$]->$[(-2+sqrt(68))/16]=0.3904$. Tilt (x,y) = ($w_{max}-w_c,w_{max}$)

## digHolo processing pipeline

The processing of a set of camera frames has several steps.

1. <u>Fourier transform the window(s) of interest in the camera frame(s).</u>

At this step, portion(s) of the camera frame that are to be Fourier transformed are selected, based on a central position (BeamCentre) on the camera around which a window of dimensions fftWindowSizeX x fftWindowSizeY will be selected.

If there are two polarisation components on the same frame, these must be on opposite sides of the width of the camera frame.

2. <u>Inverse Fourier transform the relevant off-axis window in Fourier-space, and remove offsets</u>

This is similar to the previous step, except we are now in the Fourier plane. A window is selected in Fourier-space based on the specified FourierWindowRadius and Fourier position, tilt (x,y).

This window will be inverse Fourier transformed and angle offset of the Reference field is removed. Additionally, a tilt which is related to the BeamCentre(x,y) position is applied to the window before the inverse Fourier transform is performed, such that the reconstructed field is centred at (0,0).

3. <u>Perform modal decomposition (optional)</u>

After the field has been reconstructed, there is an optional step to generate a modal representation of the field.

That is, rather than having the field represented as the summation of an array of pixels, where each pixel has it's own amplitude and phase, the field can be represented as a summation of Hermite-Gaussian modal components, or some other related basis.

Internally, the only modal basis natively supported is the Hermite-Gaussian basis. However, the user can specify their own custom matrix transformation, which maps the Hermite-Gaussian basis, onto their desired representation.

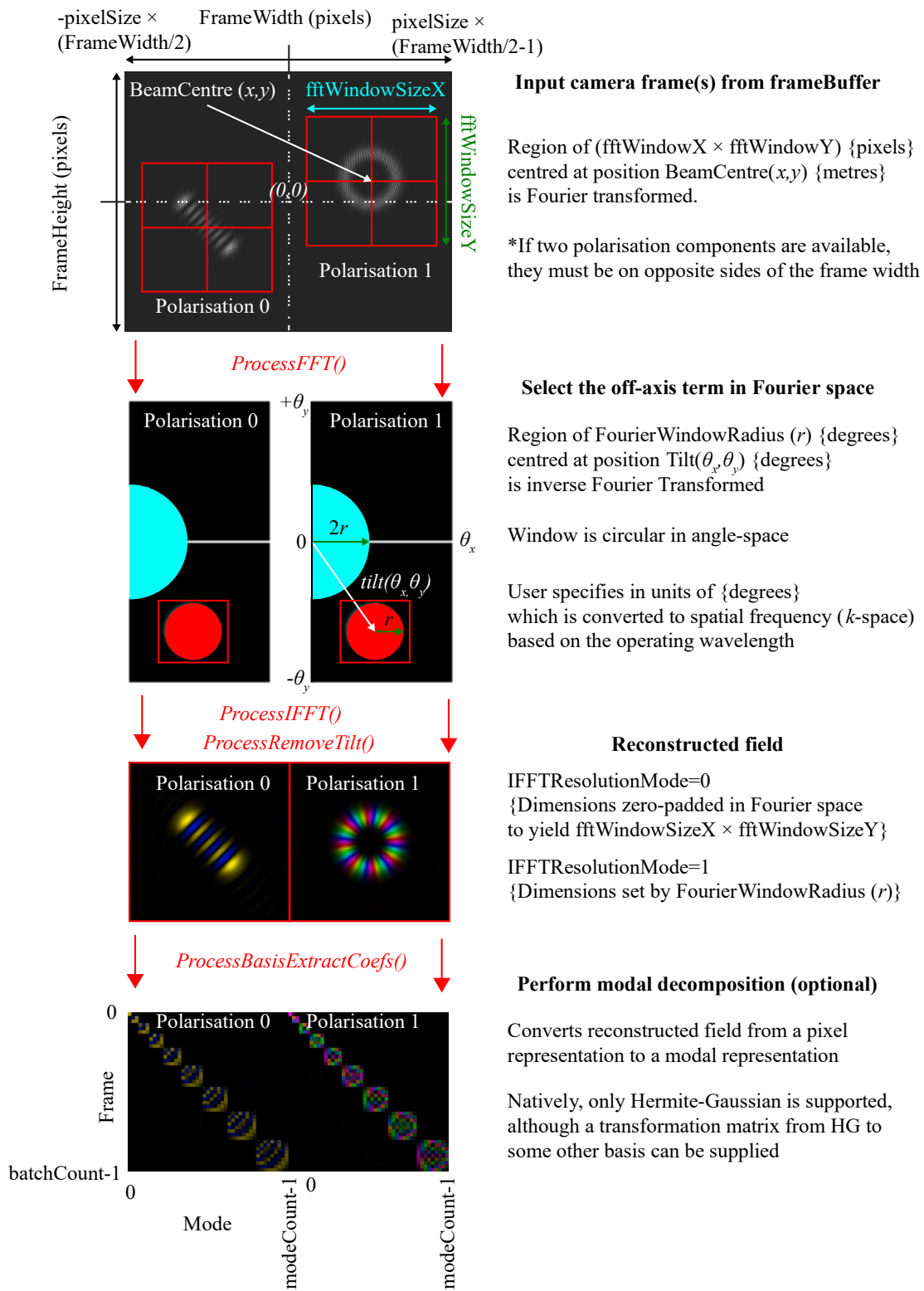

# Installation and setup

## System Requirements

CPU : x86-64 with AVX2 and FMA3 support (processors after ~2015)

## Dependencies

**FFTW 3**

https://www.fftw.org/ (Used for all FFTs and DCTs).

**Intel MKL (or other BLAS/LAPACK implementation)**

Specifically, the functions cgesvd, sgels, cgemv, cgemm.

*Linking against openBLAS https://www.openblas.net/ has also been tested. Comment out "#define MKL_ENABLE" in digHolo.cpp*

**SVML (optional)**

Trigonometric functions will be implemented using SVML if available (e.g. intel compiler and Microsoft compiler > VS2019). Otherwise, will default to hand-coded vectorised fast-math implementations.

## Compilation

The library can be compiled either as a dll/shared object, or as an executable.

The executable can be called as 'digHolo.exe {settingsFile}', where {SettingsFile} is the filename of a tab-delimited text file containing configuration information.

The examples given link against FFTW and Intel MKL.

digHolo is written in C++11 and requires support for AVX2 and FMA3 instruction sets.

**Linux (gcc or icc)**

*Shared Object*

```
g++ -std=c++11 -O3 -mavx2 -mfma -fPIC -shared digHolo.cpp -o libdigholo.so
```

*Executable*

```
g++ -std=c++11 -O3 -mavx2 -mfma digHolo.cpp -o digHolo.exe -lfftw3f -lfftw3f_threads -lmkl_intel_lp64 -lmkl_sequential -lmkl_core -lpthread -lm -ldl
```

**Windows**

Example Visual Studio solution files are provided.

*Notes on linking*

*FFTW*

Warning : always link FFTW3 first. Intel MKL also supports an FFTW interface, but does not support DCTs.

When linking statically with FFTW3, you'll have to comment out the line "#define FFTW_DLL" If you're dynamically linking, you'll have to include "#define FFTW_DLL" Because Intel MKL also includes an FFTW compatible interface. Make sure you include the FFTW library *before* MKL.

e.g.

libfftw3f-3.lib;mkl_rt.lib

NOT

mkl_rt.lib;libfftw3f-3.lib;

Otherwise, when you link, you'll actually link with the MKL FFTs, not the FFTW FFTs. Which would be fine, except for the fact that real-to-real transforms are not implemented in MKL. Real-to-real transforms are used during the 'AutoAlign' routine.

*Intel MKL*

Consult 'MKL Link link advisor' for assistance selecting the correct .libs

https://software.intel.com/content/www/us/en/develop/tools/oneapi/components/onemkl/link-line-advisor.html

Windows examples

e.g. static linked...

mkl_intel_lp64.lib;mkl_intel_thread.lib;mkl_core.lib;libiomp5md.lib

or if you're having issues with libiomp5md.lib (openMP threads), link with the sequential version

mkl_intel_lp64.lib;mkl_sequential.lib;mkl_core.lib

e.g. dynamic linked...

mkl_rt.lib

Most of your dlls will likely be in this folder, or something similar

C:\Program Files (x86)\Intel\oneAPI\mkl\latest\redist\intel64\..., if dynamically linked, you'll need to copy those into the same folder as your executable.

If linking against openMP threads library (libiomp5md.dll), don't forget the relevant dll will be in a different folder to most of the MKL dlls

C:\Program Files (x86)\Intel\oneAPI\compiler\latest\windows\redist\intel64_win\compiler\libiomp5md.dll

# Command-line execution

Although it is expected that this library would typically be compiled as a shared object/DLL, it is also possible to compile as an executable and process frames from the command-line.

In this mode of operation, the executable is called with one or more arguments specifying the location of a tab-delimited text file which specifies the relevant configuration settings.

For example,

```
digHolo.exe digholoSettings.txt
```

An example digHoloSettings.txt file is provided for reference. Camera frame data is fed in as a binary file, from a filename specified within the tab-delimited text file.

For the most part, the text file specifies which 'Set' routines to run and what values to use.

For example, to set the 'fftWindowSizeX' property to 128 pixels, the following tab-delimited line would be included.

```
fftWindowSizeX 128
```

Which would in turn would mean the following function call is invoked before the digHolo processing pipeline is run...

```
digHoloConfigSetfftWindowSizeX(handleIdx, 128);
```

Similarly, to specify a BeamCentreX for two different polarisation components of 100e-6 and 200e-6 respectively

```
BeamCentreX 100e-6 200e-6
```

Output files can also be specified, which will either be plain text or binary files depending on the type of output

For example,

```
OutputFileSummary summary.txt
OutputFilenameFields fields.bin
```

See the 'main' and 'digHoloRunBatchFromConfigFile' routines for the code itself which processes the command-line usage.

An example digHoloSettings.txt file is provided for reference.

# Benchmarks

Benchmarks can be run using the Matlab example benchmark script.

Each benchmark sweeps the dimension of the FFT window, and chooses the corresponding maximum (non-wrapping) IFFT window size.

Both single and dual polarisation are benchmarked.

Three types of benchmarks are performed...

(a) Analyses the rate at which frames can be processed from start to finish on a per frame basis as the size of the batch is increased. This benchmark has basisGroupCount=0. That is, no HG modal decomposition.

(b) Analyses the rate at which the HG modal decomposition routine can extract modal coefficients from the reconstructed fields, for a batchCount of 1. That is, when coefficients are extracted one frame at a time.

(c) The same as (b), but for a batchCount of 1000. Modes per second are the number of modal cofficients extracted per frame per mode. e.g. if extracting 1000 coefficients from each of 1000 frames took 1 second, that would be 1,000,000 modes per second.

These are simply illustrative benchmarks and not always realistic test cases. e.g. an IFFT window of size 32 x 32 = 1024 pixels, could not have 1e6 HG modes meaningfully extracted from such a low resolution.

(a) Field reconstruction (No modal decomposition)

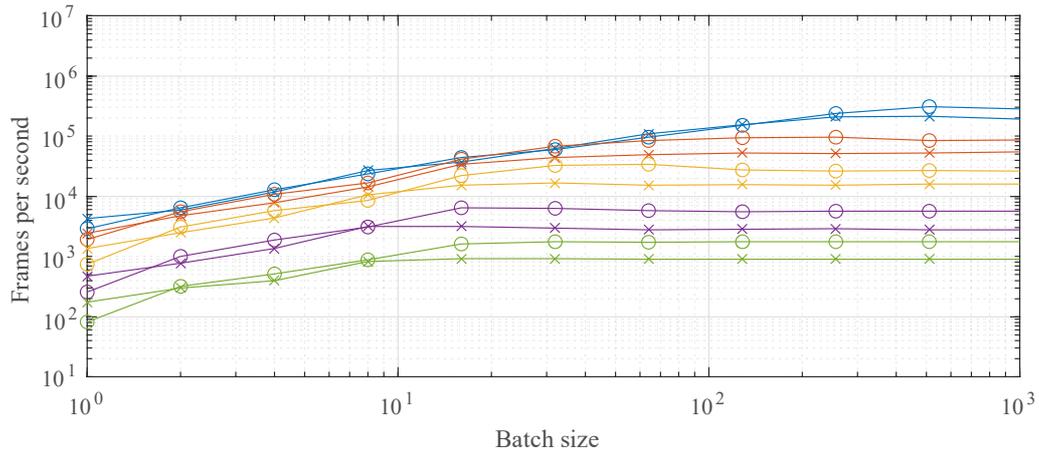

(b) Modal decomposition (BatchCount=1)

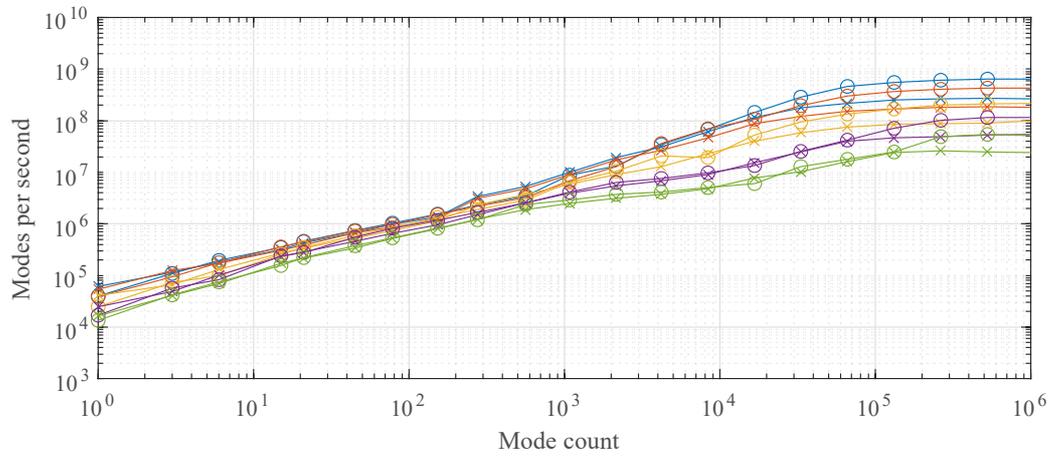

(c) Modal decomposition (BatchCount=1000)

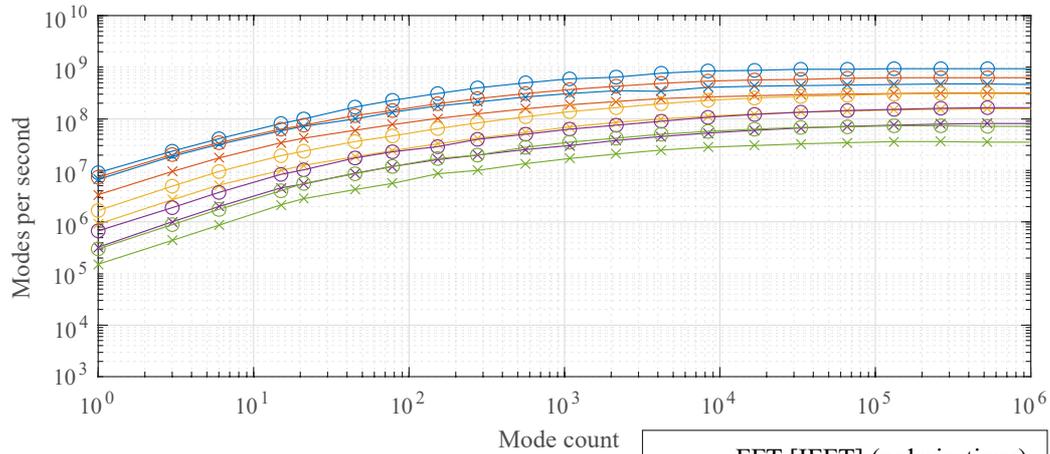

CPU : Intel Core i9 7980XE (18 @ 2.6 GHz) [2017]
Threads : 36
FFTW planning : Exhaustive
RAM : DDR4 3200 MHz Quad Channel
L1 cache : 18 × 64 kB
L2 cache : 18 × 1 MB
L3 cache : 24.75 MB

| FFT [IFFT] (polarisations) |
| --- |
| 64 × 64 [32 × 32] (1 × pol) |
| 64 × 64 [32 × 32] (2 × pol) |
| 128 × 128 [48 × 48] (1 × pol) |
| 128 × 128 [48 × 48] (2 × pol) |
| 256 × 256 [96 × 96] (1 × pol) |
| 256 × 256 [96 × 96] (2 × pol) |
| 512 × 512 [176 × 176] (1 × pol) |
| 512 × 512 [176 × 176] (2 × pol) |
| 1024 × 1024 [336 × 336] (1 × pol) |
| 1024 × 1024 [336 × 336] (2 × pol) |

(a) Field reconstruction (No modal decomposition)

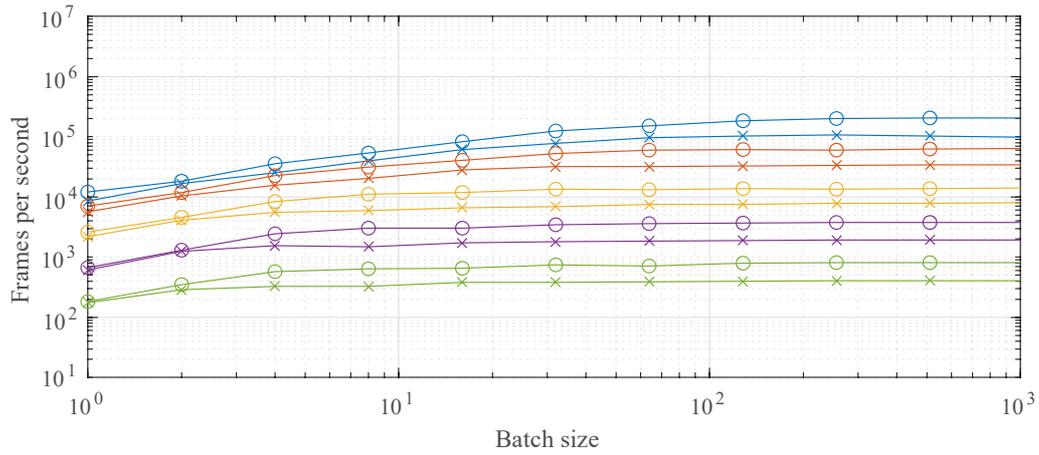

(b) Modal decomposition (BatchCount=1)

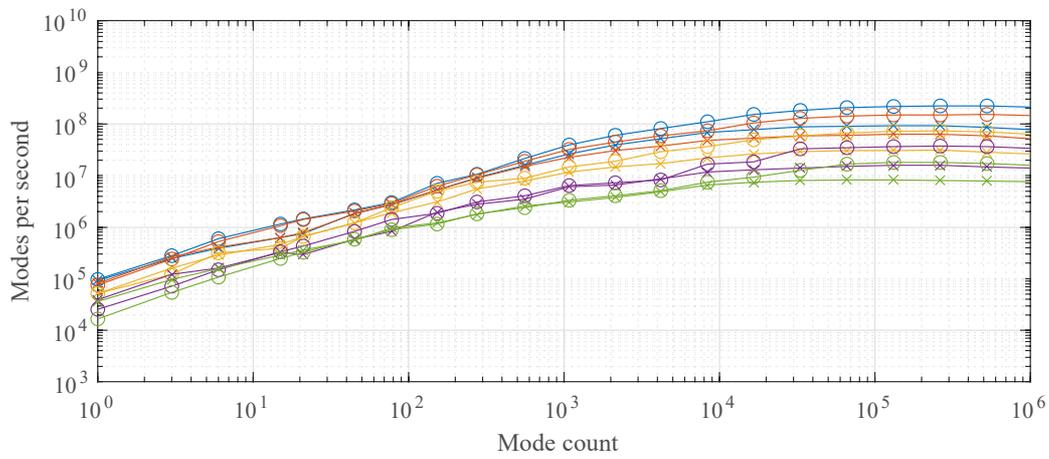

(c) Modal decomposition (BatchCount=1000)

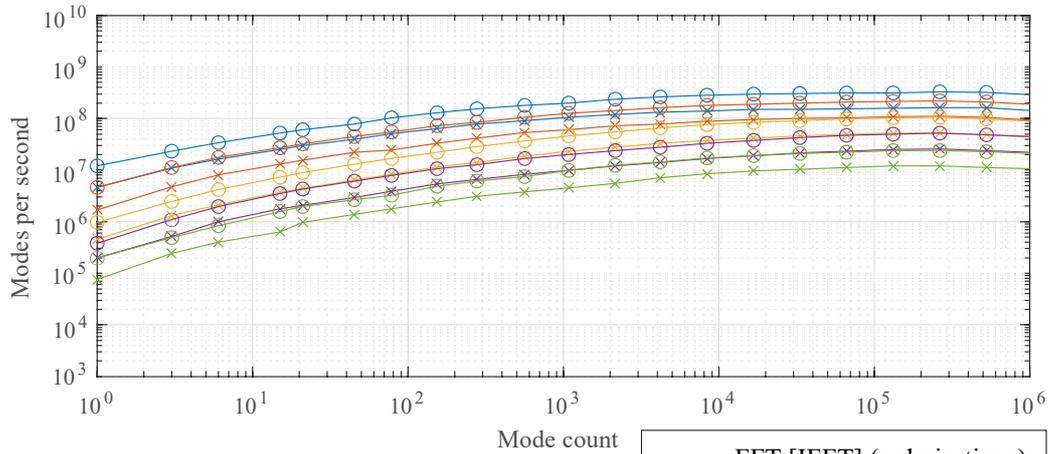

CPU : Intel Core i7 5820K (6 @ 3.3 GHz) [2014]
Threads : 12
FFTW planning : Exhaustive
RAM : DDR4 2166 MHz Quad Channel
L1 cache : 6 × 64 kB
L2 cache : 6 × 256 kB
L3 cache : 15.00 MB

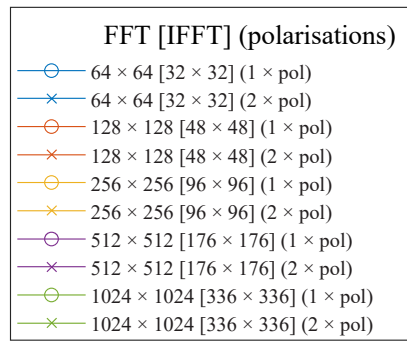

# Module Index

## Modules

Here is a list of all modules:

# Module Documentation

## Error codes

Error codes returned by functions.

### Macros

- #define DIGHOLO_ERROR_SUCCESS   0
- #define DIGHOLO_ERROR_ERROR   1
- #define DIGHOLO_ERROR_INVALIDHANDLE   2
- #define DIGHOLO_ERROR_NULLPOINTER   3
- #define DIGHOLO_ERROR_SETFRAMEBUFFERDISABLED   4
- #define DIGHOLO_ERROR_INVALIDDIMENSION   5
- #define DIGHOLO_ERROR_INVALIDPOLARISATION   6
- #define DIGHOLO_ERROR_INVALIDAXIS   7
- #define DIGHOLO_ERROR_INVALIDARGUMENT   8
- #define DIGHOLO_ERROR_MEMORYALLOCATION   9
- #define DIGHOLO_ERROR_FILENOTCREATED   10
- #define DIGHOLO_ERROR_FILENOTFOUND   11

### Detailed Description

Error codes returned by functions.

### Macro Definition Documentation

#### #define DIGHOLO_ERROR_ERROR   1

Generic error

#### #define DIGHOLO_ERROR_FILENOTCREATED   10

File could not be created. (e.g. trying to redirect the console to a file path that doesn't exist)

#### #define DIGHOLO_ERROR_FILENOTFOUND   11

File could not be found/opened. (e.g. specifying a settings file path to open that doesn't exist)

#### #define DIGHOLO_ERROR_INVALIDARGUMENT   8

A specified argument is not valid. (e.g. attempt to set value to out of range or otherwise meaningless value)

#### #define DIGHOLO_ERROR_INVALIDAXIS   7

Specified axis doesn't exist. (e.g. attempt to address third axis of a 2D system, or a negative axis)

**#define DIGHOLO_ERROR_INVALIDDIMENSION   5**

    Dimension provided is invalid. (e.g. less than zero or greater than max possible value)

**#define DIGHOLO_ERROR_INVALIDHANDLE   2**

    handleIdx provided doesn't exist

**#define DIGHOLO_ERROR_INVALIDPOLARISATION   6**

    Specified polarisation component doesn't exist. (e.g. there is only 1 polarisation component, but attempt to address second component)

**#define DIGHOLO_ERROR_MEMORYALLOCATION   9**

    Memory allocation failed

**#define DIGHOLO_ERROR_NULLPOINTER   3**

    A pointer provided was null

**#define DIGHOLO_ERROR_SETFRAMEBUFFERDISABLED   4**

    Requested to update the frame buffer, but it is currently disabled for update. Deprecated code.

**#define DIGHOLO_ERROR_SUCCESS   0**

    Success (no error)

## units

Working units of the library.

### Macros

- #define DIGHOLO_UNIT_PIXEL   1
- #define DIGHOLO_UNIT_LAMBDA   1
- #define DIGHOLO_UNIT_ANGLE   (pi/180.0f)

### Detailed Description

Working units of the library.

### Macro Definition Documentation

#### #define DIGHOLO_UNIT_ANGLE   (pi/180.0f)

Angle units (e.g. tilt and FourierWindowRadius) (1 = radian, pi/180 = degrees)

#### #define DIGHOLO_UNIT_LAMBDA   1

Wavelengths units (1 = metres)

#### #define DIGHOLO_UNIT_PIXEL   1

Pixels and camera plane positions in units (1 = metres)

# Creating and destroying digHoloObjects

Functions for creating and destroying digHoloObjects.

## Functions

- int digHoloCreate ()
  *Creation of a new digHoloObject.*

- int digHoloDestroy (int handleIdx)
  *Deallocates memory and terminates threads associated with this digHoloObject.*

## Detailed Description

Functions for creating and destroying digHoloObjects.

## Function Documentation

### int digHoloCreate ()

Creation of a new digHoloObject.

Creates a default digHoloObject and adds it to the internal object array. Implements some initialisation such as creating a threadpool for the digHoloObject. External code calling this library will pass the handleIdx returned by this function to almost all future functions.

**Returns**

integer which uniquely identifies the newly created object, and is used by future functions to indicate which digHoloObject it applies to.

### int digHoloDestroy (int *handleIdx*)

Deallocates memory and terminates threads associated with this digHoloObject.

**Parameters**

| | |
|---|---|
| *handleIdx* | is the enumeration of the digHoloObject to be destroyed. |

## off-axis digital holography configuration

Configuring the source and dimensions of frame data.

### Modules

- Frame buffer
  *Configuring the source and dimensions of frame data.*
- Batch processing
  *Configuring batch processing of frames.*
- Window sizes
  *Configuring the size of the windows in the frame plane and Fourier plane, as well as wavelength of operation.*
- Wavelength
  *Configuring the current operating wavelength. (e.g. for converting angles to k-space)*
- Alignment
  *Configuring key-properties such as reference beam tilt, defocus, beam centre, etc.)*
- Modal decomposition
  *Routines for configuring the spatial modal basis (if any) (e.g. Hermite-Gaussian, Laguerre-Gaussian)*

### Detailed Description

Configuring the source and dimensions of frame data.

# Frame buffer

Configuring the source and dimensions of frame data.

## Functions

- int digHoloSetFrameBuffer (int handleIdx, float *buffer)
  *Sets the internal frame buffer pointer to an external buffer containing potentially multiple camera frames worth of pixels in float32 format.*

- float * digHoloGetFrameBuffer (int handleIdx)
  *Returns the pointer to the float32 frame buffer.*

- int digHoloSetFrameBufferUint16 (int handleIdx, unsigned short *buffer, int transposeMode)
  *Sets the internal frame buffer pointer to an external buffer containing potentially multiple camera frames worth of pixels in uint16 format.*

- unsigned short * digHoloGetFrameBufferUint16 (int handleIdx, int *transposeMode)
  *Returns the pointer to the uint16 frame buffer.*

- int digHoloSetFrameBufferFromFile (int handleIdx, const char *fname)
  *Creates and sets an internal frame buffer by reading in uint16 pixel data from a binary file.*

- int digHoloConfigSetFrameDimensions (int handleIdx, int width, int height)
  *Sets the full dimensions of the frames in the frame buffer. (e.g. width,height = 640x512)*

- int digHoloConfigGetFrameDimensions (int handleIdx, int *width, int *height)
  *Returns the current full dimensions of the frames in the frame buffer. (e.g. width,height = 640x512)*

- int digHoloConfigSetFramePixelSize (int handleIdx, float pixelSize)
  *Sets the current pixel width/height/pitch of the frame pixels.*

- float digHoloConfigGetFramePixelSize (int handleIdx)
  *Returns the current pixel width/height/pitch of the frame pixels.*

- int digHoloConfigSetPolCount (int handleIdx, int polCount)
  *Sets the number of polarisation components per frame (1 or 2)*

- int digHoloConfigGetPolCount (int handleIdx)
  *Gets the number of polarisation components per frame (1 or 2).*

- int digHoloConfigSetRefCalibrationIntensity (int handleIdx, unsigned short *cal, int wavelengthCount, int width, int height)
  *Define the intensity of the Reference wave, which will be used to calibrate out a non-uniform Reference wave.*

- int digHoloConfigSetRefCalibrationField (int handleIdx, complex64 *cal, int wavelengthCount, int width, int height)

*Define the field of the Reference wave, which will be used to calibrate out a non-uniform and aberrated Reference wave.*

- int digHoloConfigSetRefCalibrationFromFile (int handleIdx, const char *fname, int wavelengthCount, int width, int height)
  *Define the Reference wave from file.*

- int digHoloConfigSetRefCalibrationEnabled (int handleIdx, int enabled)
  *Defines whether a previously set Reference wave calibration should be used or not.*

- int digHoloConfigGetRefCalibrationEnabled (int handleIdx)
  *Returns whether Reference wave calibration is currently enabled or disabled.*

- complex64 * digHoloConfigGetRefCalibrationFields (int handleIdx, int *wavelengthCount, int *polCount, float **x, float **y, int *width, int *height)
  *Return Reference wave calibration and corresponding x and y axis.*

- int digHoloConfigSetFillFactorCorrectionEnabled (int handleIdx, int enabled)
  *Defines whether the effect of the finite dimensions of the camera pixels should be calibrated out.*

- int digHoloConfigGetFillFactorCorrectionEnabled (int handleIdx)
  *Returns whether the effect of the finite dimensions of the camera pixels should be calibrated out.*

## Detailed Description

Configuring the source and dimensions of frame data.

## Function Documentation

### int digHoloConfigGetFillFactorCorrectionEnabled (int *handleIdx*)

Returns whether the effect of the finite dimensions of the camera pixels should be calibrated out.

When enabled, the camera frames are assumed to have a 100% fill factor, and the corresponding sinc envelope is calibrated out from the Fourier plane. This calibration is performed during the IFFT process.

**Parameters**

| in | *handleIdx* | : enumerated handle index of the digHoloObject |
|---|---|---|

**Returns**
   enabled : Enable (1) or disable (0) fill factor correction in the Fourier plane.

### int digHoloConfigGetFrameDimensions (int *handleIdx*, int * *width*, int * *height*)

Returns the current full dimensions of the frames in the frame buffer. (e.g. width,height = 640x512)

**Parameters**

| in | *handleIdx* | : enumerated handle index of the digHoloObject |
|---|---|---|
| out | *width* | : width of the frames. Typically width is the longer dimension on a regular camera. The x-axis. |
| out | *height* | : height of the frames. Typically the height is the shorter dimension on a regular camera. The y-axis. |

**Returns**

errorCode : [DIGHOLO_ERROR_SUCCESS, DIGHOLO_ERROR_INVALIDHANDLE]

## float digHoloConfigGetFramePixelSize (int *handleIdx*)

Returns the current pixel width/height/pitch of the frame pixels.

**Parameters**

| in | *handleIdx* | : enumerated handle index of the digHoloObject |
|---|---|---|

**Returns**

pixelSize : width=height=pitch of frame pixel. Will return 0 for invalid handle index.

## int digHoloConfigGetPolCount (int *handleIdx*)

Gets the number of polarisation components per frame (1 or 2).

Defines the number of FFTs that will be performed per frame. polCount windows of the same dimensions are Fourier transformed.

**Parameters**

| in | *handleIdx* | : enumerated handle index of the digHoloObject |
|---|---|---|

**Returns**

polCount : [1 2]. Returns zero if invalid handlex index is supplied.

## int digHoloConfigGetRefCalibrationEnabled (int *handleIdx*)

Returns whether Reference wave calibration is currently enabled or disabled.

**Parameters**

| in | *handleIdx* | : enumerated handle index of the digHoloObject |
|---|---|---|

**Returns**

enabled : Enable (1) or disable (0) the use of Reference wave calibration. Returns zero for an invalid handle.

## complex64 * digHoloConfigGetRefCalibrationFields (int *handleIdx*, int * *wavelengthCount*, int * *polCount*, float ** *x*, float ** *y*, int * *width*, int * *height*)

Return Reference wave calibration and corresponding x and y axis.

Returns a pointer to the final applied Reference wave calibration field, which will have the same dimensions as the reconstructed field. In general, this is not the same raw calibration source set by the user in digHoloConfigSetRefCalibration. This is the final calibration that was applied to the reconstructed fields. Compared to the full-frame calibration source specified by the user. These fields will be of different dimensions (due to windowing in the camera plane and Fourier plane during processing), will have conjugate phase and inverse amplitude. This function follows the same calling convention as routines such as 'digHoloGetFields()'. The Reference wave calibration will only be available after digHolo processing has been performed, as the Reference wave calibration must also be processed to match the dimensions and properties of the reconstructed fields.

**Parameters**

| in | *handleIdx* | : enumerated handle index of the digHoloObject |
|---|---|---|
| out | *wavelengthCount* | : pointer to an int32 where the number of fields in the calibration per polarisation will be returned. |
| out | *polCount* | : pointer to an int32 where the number of polarisation components will be returned. |
| out | *x* | : pointer to an array containing the x-axis of the field. |
| out | *y* | : pointer to an array containing the y-axis of the field. |
| out | *width* | : pointer to an int32 where the length of the x-axis of the field per polarisation will be returned. |
| out | *height* | : pointer to an int32 where the length of the y-axis of the field per polarisation will be returned. |

**Returns**

fields : wavelengthCount x polCount x width x height complex64 array containing the fields

### int digHoloConfigSetFillFactorCorrectionEnabled (int *handleIdx*, int *enabled*)

Defines whether the effect of the finite dimensions of the camera pixels should be calibrated out.

When enabled, the camera frames are assumed to have a 100% fill factor, and the corresponding sinc envelope is calibrated out from the Fourier plane. This calibration is performed during the IFFT process.

**Parameters**

| in | *handleIdx* | : enumerated handle index of the digHoloObject |
|---|---|---|
| in | *enabled* | : Enable (1) or disable (0) fill factor correction in the Fourier plane. |

**Returns**

errorCode : [DIGHOLO_ERROR_SUCCESS DIGHOLO_ERROR_INVALIDHANDLE]

### int digHoloConfigSetFrameDimensions (int *handleIdx*, int *width*, int *height*)

Sets the full dimensions of the frames in the frame buffer. (e.g. width,height = 640x512)

**Parameters**

| in | *handleIdx* | : enumerated handle index of the digHoloObject |
|---|---|---|
| in | *width* | : width of the frames. Typically width is the longer dimension on a regular camera. The x-axis. Must be multiple of DIGHOLO_PIXEL_QUANTA. |
| in | *height* | : height of the frames. Typically the height is the shorter dimension on a regular camera. The y-axis. Must be multiple of DIGHOLO_PIXEL_QUANTA. |

**Returns**

errorCode : [DIGHOLO_ERROR_SUCCESS, DIGHOLO_ERROR_INVALIDHANDLE]

### int digHoloConfigSetFramePixelSize (int *handleIdx*, float *pixelSize*)

Sets the current pixel width/height/pitch of the frame pixels.

**Parameters**

| | | |
|---|---|---|
| in | *handleIdx* | : enumerated handle index of the digHoloObject |
| in | *pixelSize* | : width=height=pitch of frame pixel |

**Returns**

    errorCode : [DIGHOLO_ERROR_SUCCESS DIGHOLO_ERROR_INVALIDHANDLE DIGHOLO_ERROR_INVALIDDIMENSION]

### int digHoloConfigSetPolCount (int *handleIdx*, int *polCount*)

Sets the number of polarisation components per frame (1 or 2)

**Parameters**

| | | |
|---|---|---|
| in | *handleIdx* | : enumerated handle index of the digHoloObject |
| in | *polCount* | : polarisation components per frame (1 or 2) |

**Returns**

    errorCode : [DIGHOLO_ERROR_SUCCESS DIGHOLO_ERROR_INVALIDHANDLE DIGHOLO_ERROR_INVALIDPOLARISATION]

### int digHoloConfigSetRefCalibrationEnabled (int *handleIdx*, int *enabled*)

Defines whether a previously set Reference wave calibration should be used or not.

**Parameters**

| | | |
|---|---|---|
| in | *handleIdx* | : enumerated handle index of the digHoloObject |
| in | *enabled* | : Enable (1) or disable (0) the use of Reference wave calibration. |

**Returns**

    errorCode : [DIGHOLO_ERROR_SUCCESS DIGHOLO_ERROR_INVALIDHANDLE]

### int digHoloConfigSetRefCalibrationField (int *handleIdx*, complex64 * *cal*, int *wavelengthCount*, int *width*, int *height*)

Define the field of the Reference wave, which will be used to calibrate out a non-uniform and aberrated Reference wave.

See 'digHoloConfigSetRefCalibrationIntensity' for further details. This routine is similar to digHoloSetRefCalibrationIntensity, except the user specifies the Reference wave as a full field, and hence can compensate not only for non-uniform intensity in the Reference wave, but also aberrations.

**Parameters**

| | | |
|---|---|---|
| in | *handleIdx* | : enumerated handle index of the digHoloObject |
| in | *cal* | : An array of pixels (wavelengthCount x width x height) specifying the field of the Reference wave |
| in | *wavelengthCount* | : The number of frames in 'cal' which is assumed to specify Reference waves captured at different wavelengths matching those of the FrameBuffer. |
| in | *width* | : Width of a single frame (should be consistent with digHoloFrameWidth) |
| in | *height* | : Height of a single frame (should be consistent with |

|   |   |   | digHoloFrameHeight) |
|---|---|---|---|

**Returns**

errorCode : [DIGHOLO_ERROR_SUCCESS DIGHOLO_ERROR_INVALIDHANDLE DIGHOLO_ERROR_NULLPOINTER DIGHOLOERROR_INVALIDARGUMENT]

### int digHoloConfigSetRefCalibrationFromFile (int *handleIdx*, const char * *fname*, int *wavelengthCount*, int *width*, int *height*)

Define the Reference wave from file.

See 'digHoloConfigSetRefCalibrationIntensity' for further details. This routine will load a binary file containing the array 'cal' from the digHoloConfigSetRefCalibrationIntensity or digHoloConfigSetRefCalibrationField routines. Based on the specified width, height and wavelengthCount, this routine will infer from the filesize whether the user has specified an Intensity in uint16 format, or a field in 2xfloat32 complex format.

**Parameters**

| in | *handleIdx* | : enumerated handle index of the digHoloObject |
|---|---|---|
| in | *fname* | : Filename of a binary file containing the Reference wave as either an intensity (uint16) or a field (2xfloat32). |
| in | *wavelengthCount* | : The number of frames in 'cal' which is assumed to specify Reference waves captured at different wavelengths matching those of the FrameBuffer. |
| in | *width* | : Width of a single frame (should be consistent with digHoloFrameWidth) |
| in | *height* | : Height of a single frame (should be consistent with digHoloFrameHeight) |

**Returns**

errorCode : [DIGHOLO_ERROR_SUCCESS DIGHOLO_ERROR_FILENOTFOUND DIGHOLO_ERROR_INVALIDHANDLE DIGHOLO_ERROR_NULLPOINTER DIGHOLOERROR_INVALIDARGUMENT]

### int digHoloConfigSetRefCalibrationIntensity (int *handleIdx*, unsigned short * *cal*, int *wavelengthCount*, int *width*, int *height*)

Define the intensity of the Reference wave, which will be used to calibrate out a non-uniform Reference wave.

Specifies the intensity of the reference wave. Used to compensate for non-uniform Reference wave. The user inputs wavelengthCount frames of dimensions width x height, which should match digHoloFrameWidth and digHoloFrameHeight. These frames would be the measured intensity of the Reference wave alone, and can be specified in a wavelength-dependent fashion (wavelengthCount>1). If wavelengthCount==1, the same calibration is used for all wavelengths. It is assumed that wavelength-dependent Reference wave calibrations match the wavelengths specified using routines such as digHoloConfigSetWavelengths. The calibration is read in only when this function is called. That is, if 'cal' is later updated externally, this will not be reflected in the calibration applied. Similarly, the pointer 'cal' can be destroyed after this function is called. The maximum intensity value over all pixels (wavelengthCount x width x height) is found, and then the calibration is normalised relative to this value. Hence it does not matter what the absolute values of the pixels in 'cal' are, and the Reference wave can be measured on the camera using whatever exposure and power level is convenient. That is, the power of the Reference wave in the calibration, need not be the same as will be used for later digital holography processing. When working with dual-polarisation, each polarisation should be non-overlapping within the frame, as the calibration is a function of camera pixel position and is not specified independently for each polarisation component.

The RefCalibration is automatically enabled when this function is called. To later disable the calibration use digHoloConfigSetRefCalibrationEnabled(). Calling this function with an invalid argument (e.g. null pointer, width<=0, height<=0 and/or wavelengthCount<=0) will disable calibration. i.e. Same effect as digHoloConfigSetRefCalibrationEnabled(handleIdx,false))

**Parameters**

| in | *handleIdx* | : enumerated handle index of the digHoloObject |
|---|---|---|
| in | *cal* | : An array of pixels (wavelengthCount x width x height) specifying the intensity of the Reference wave |
| in | *wavelengthCount* | : The number of frames in 'cal' which is assumed to specify Reference waves captured at different wavelengths matching those of the FrameBuffer. |
| in | *width* | : Width of a single frame (should be consistent with digHoloFrameWidth) |
| in | *height* | : Height of a single frame (should be consistent with digHoloFrameHeight) |

**Returns**

errorCode : [DIGHOLO_ERROR_SUCCESS DIGHOLO_ERROR_INVALIDHANDLE DIGHOLO_ERROR_NULLPOINTER DIGHOLOERROR_INVALIDARGUMENT]

**float * digHoloGetFrameBuffer (int    *handleIdx*)**

Returns the pointer to the float32 frame buffer.

If operating from a float32 frame buffer, this pointer will be the same as that set by digHoloSetFrameBuffer() or similar routines such as digHoloSetBatch() If operating from an uint16 frame buffer, this pointer will be an internally buffer allocated and populated soon before the internal FFT is invoked.

**Parameters**

| in | *handleIdx* | : enumerated handle index of the digHoloObject |
|---|---|---|

**Returns**

pointer : pointer to either an internal (uint16 mode), or external (float32 mode) buffer, or a null pointer if the handleIdx is invalid or the buffer is not set.

**unsigned short * digHoloGetFrameBufferUint16 (int    *handleIdx*, int *    *transposeMode*)**

Returns the pointer to the uint16 frame buffer.

**Parameters**

| in | *handleIdx* | : enumerated handle index of the digHoloObject |
|---|---|---|
| out | *transposeMode* | : returns the flag which indicates whether the frames should be transposed during the uint16->float32 conversion process. |

**Returns**

pointer : pointer to either uint16 frame buffer, or a null pointer if the handleIdx is invalid or the buffer is not set.

**int digHoloSetFrameBuffer (int    *handleIdx*, float *    *buffer*)**

Sets the internal frame buffer pointer to an external buffer containing potentially multiple camera frames worth of pixels in float32 format.

Similar to digHoloSetBatch() below, and typically digHoloSetBatch() would be preferred/safer usage. digHoloSetBatch() sets SetFrameBufferEnabled = false, and it stays false until you run 'Resume'. This is to prevent fights between GUI threads updating viewports, and routines like digHoloSetBatch trying to process specific frames. This aspect using of using SetFrameBufferEnabled should be deprecated.

**Parameters**

| in | *handleIdx* | : enumerated handle index of the digHoloObject |
|---|---|---|
| in | *buffer* | : pointer to a buffer of pixel data in float32 format, corresponding with potentially multiple camera frames. |

**Returns**

errorCode : [DIGHOLO_ERROR_SUCCESS, DIGHOLO_ERROR_INVALIDHANDLE, DIGHOLO_ERROR_NULLPOINTER]

## int digHoloSetFrameBufferFromFile (int *handleIdx*, const char * *fname*)

Creates and sets an internal frame buffer by reading in uint16 pixel data from a binary file.

Pixel data can be read in from file, without having to immediately specify the frameWidth, frameHeight and frameCount. However before the data can actually be used as a source of frames, this will need to be specified using routines like digHoloConfigSetFrameDimensions().

**Parameters**

| in | *handleIdx* | : enumerated handle index of the digHoloObject |
|---|---|---|
| in | *fname* | : file path to a binary file containing frame(s) of pixel data in uint16 format. |

**Returns**

errorCode : [DIGHOLO_ERROR_SUCCESS, DIGHOLO_ERROR_INVALIDHANDLE, DIGHOLO_ERROR_INVALIDDIMENSION, DIGHOLO_ERROR_FILENOTFOUND, DIGHOLO_ERROR_MEMORYALLOCATION]

## int digHoloSetFrameBufferUint16 (int *handleIdx*, unsigned short * *buffer*, int *transposeMode*)

Sets the internal frame buffer pointer to an external buffer containing potentially multiple camera frames worth of pixels in uint16 format.

The equivalent float32 routine (digHoloSetFrameBuffer) is the preferred method of inputting frame data as there is a ~20%? performance penalty due to the required conversion from uint16->float32. As opposed to float32 data, which can be fed directly into the FFT, uint16 data must be converted first, and as the buffer is external, the digHoloObject has no way of knowing if the data has changed since the last conversion, and hence must reconvert. Unlike float32. Calling digHoloSetFrameBuffer() will reset the frame source as a float32 buffer.

**Parameters**

| in | *handleIdx* | : enumerated handle index of the digHoloObject |
|---|---|---|
| in | *buffer* | : pointer to a buffer of pixel data in uint16 format, corresponding with potentially multiple camera frames. |
| in | *transposeMode* | : setting this flag to true will transpose the individual camera frames during the uint16->float32 format. |

**Returns**

errorCode : [DIGHOLO_ERROR_SUCCESS, DIGHOLO_ERROR_INVALIDHANDLE, DIGHOLO_ERROR_NULLPOINTER]

# Batch processing

Configuring batch processing of frames.

## Modules

- Averaging modes
  *The reconstructed fields can have averaging applied, whereby multiple frames are constructively averaged to produce a reconstructed field with less noise.*

## Functions

- int digHoloSetBatch (int handleIdx, int batchCount, float *frameBuffer)
  *Sets the parameters for processing a batch of batchCount frames stored in a float32 frameBuffer.*

- int digHoloSetBatchAvg (int handleIdx, int batchCount, float *frameBuffer, int avgCount, int avgMode)
  *Sets the parameters for processing a batch of (batchCount x avgCount) frames stored in a float32 frameBuffer, with averaging.*

- int digHoloSetBatchUint16 (int handleIdx, int batchCount, unsigned short *frameBuffer, int transpose)
  *Sets the parameters for processing a batch of batchCount frames stored in a uint16 frameBuffer.*

- int digHoloSetBatchAvgUint16 (int handleIdx, int batchCount, unsigned short *frameBuffer, int avgCount, int avgMode, int transpose)
  *Sets the parameters for processing a batch of batchCount frames stored in a uint16 frameBuffer.*

- int digHoloConfigSetBatchCount (int handleIdx, int batchCount)
  *Sets the current batch size for processing frames.*

- int digHoloConfigGetBatchCount (int handleIdx)
  *Gets the current batch size for processing frames.*

- int digHoloConfigSetBatchAvgCount (int handleIdx, int avgCount)
  *Also see digHoloSetBatchAvg().*

- int digHoloConfigGetBatchAvgCount (int handleIdx)
- int digHoloConfigSetBatchCalibration (int handleIdx, complex64 *cal, int polCount, int batchCount)
- complex64 * digHoloConfigGetBatchCalibration (int handleIdx, int *polCount, int *batchCount)
- int digHoloConfigSetBatchCalibrationFromFile (int handleIdx, const char *fname, int polCount, int batchCount)
- int digHoloConfigSetBatchCalibrationEnabled (int handleIdx, int enabled)
- int digHoloConfigGetBatchCalibrationEnabled (int handleIdx)
- int digHoloConfigSetBatchAvgMode (int handleIdx, int avgMode)
- int digHoloConfigGetBatchAvgMode (int handleIdx)

## Detailed Description

Configuring batch processing of frames.

## Function Documentation

### int digHoloConfigGetBatchAvgCount (int *handleIdx*)

Returns the amount of averaging to perform when processing frames.

Also see digHoloSetBatchAvg().

**Parameters**

| in | *handleIdx* | : enumerated handle index of the digHoloObject |
|----|-------------|------------------------------------------------|

**Returns**

avgCount : the number of frames to be averaged per batch. Returns zero for invalid handle index.

### int digHoloConfigGetBatchAvgMode (int *handleIdx*)

Returns the way frames to be averaged are organise in the frame buffer (DIGHOLO_AVGMODE_SEQUENTIAL, DIGHOLO_AVGMODE_INTERLACED,DIGHOLO_AVGMODE_SEQUENTIALSWEEP).

Also see digHoloSetBatchAvg().

**Parameters**

| in | *handleIdx* | : enumerated handle index of the digHoloObject |
|----|-------------|------------------------------------------------|

**Returns**

avgMode : [DIGHOLO_AVGMODE_SEQUENTIAL, DIGHOLO_AVGMODE_INTERLACED,DIGHOLO_AVGMODE_SEQUENTIALSWEEP] . Returns zero for invalid handle index.

### complex64 * digHoloConfigGetBatchCalibration (int *handleIdx*, int * *polCount*, int * *batchCount*)

Returns pointer to internal array containing the batchCalibration.

See digHoloConfigSetBatchCalibration for further information.

**Parameters**

| in  | *handleIdx* | : enumerated handle index of the digHoloObject |
|-----|-------------|------------------------------------------------|
| out | *polCount*  | : Pointer to an int32 where the polCount dimension of cal will be written. |
| out | *batchCount*| : Pointer to an int32 where the batchCount dimension of cal will be written. |

**Returns**

cal : Pointer to a polCount x batchCount array of complex numbers to be multiplied to each reconstructed field within a processing batch.

### int digHoloConfigGetBatchCalibrationEnabled (int *handleIdx*)

Returns whether the batchCalibration is enabled or disabled.

See digHoloConfigSetBatchCalibration for further information.

**Parameters**

| in | *handleIdx* | : enumerated handle index of the digHoloObject |
|----|-------------|------------------------------------------------|

Returns

    enabled : Enable (1) or disable (0) the batchCalibration if available. An invalid handleIdx will return 0.

### int digHoloConfigGetBatchCount (int *handleIdx*)

Gets the current batch size for processing frames.

**Parameters**

| in | *handleIdx* | : enumerated handle index of the digHoloObject |
|---|---|---|

**Returns**

    batchCount : the number of frames to be processed. Returns zero for invalid handle index

### int digHoloConfigSetBatchAvgCount (int *handleIdx*, int *avgCount*)

Also see digHoloSetBatchAvg().

Sets the amount of averaging to perform when processing frames.

**Parameters**

| in | *handleIdx* | : enumerated handle index of the digHoloObject |
|---|---|---|
| in | *avgCount* | : the number of frames to be averaged per batch. avgCount<1 will default to 0, but will not raise an error. |

**Returns**

    errorCode : [DIGHOLO_ERROR_SUCCESS, DIGHOLO_ERROR_INVALIDHANDLE]

### int digHoloConfigSetBatchAvgMode (int *handleIdx*, int *avgMode*)

Sets the way frames to be averaged are organise in the frame buffer (DIGHOLO_AVGMODE_SEQUENTIAL, DIGHOLO_AVGMODE_INTERLACED,DIGHOLO_AVGMODE_SEQUENTIALSWEEP).

Also see digHoloSetBatchAvg().

**Parameters**

| in | *handleIdx* | : enumerated handle index of the digHoloObject |
|---|---|---|
| in | *avgMode* | : Sequential (DIGHOLO_AVGMODE_SEQUENTIAL), Interlaced (DIGHOLO_AVGMODE_INTERLACED). Sequential wavelength sweep (DIGHOLO_AVGMODE_SEQUENTIALSWEEP). Other values will default to 0. |

**Returns**

    errorCode : [DIGHOLO_ERROR_SUCCESS, DIGHOLO_ERROR_INVALIDHANDLE]

### int digHoloConfigSetBatchCalibration (int *handleIdx*, complex64 * *cal*, int *polCount*, int *batchCount*)

Defines an array polCount x batchCount of complex numbers that will be multiplied to each reconstructed field.

This array can be used to calibrate out frame dependencies within a batch. For example, if frames within a batch are captured at different camera exposure or power, this apparent difference in power could be removed using this calibration. Similarly, properties such as delay and/or dispersion could be removed from wavelength-dependent batches. The user can specify a polarisation-dependent calibration by setting polCount>1. However it is also possible to use a polarisation-independent calibration using polCount==1, even if the reconstructed fields have more than 1 polarisation. In such a case, the same calibration is applied to both polarisations. If a batch larger than batchCount is processed whilst this calibration is enabled, the applied batchIdx will wrap around by modular division (calIdx = batchIdxbatchCount) When this function is called, BatchCalibration will automatically be enabled. If this function is called with invalid arguments (e.g. null pointer, polCount<1, batchCount<1) then BatchCalibration will be automatically disabled. The array 'cal' is copied internally, and hence can be destroyed after this is called, and external updates to cal will not be seen internally without another call of this function.

**Parameters**

| in | *handleIdx* | : enumerated handle index of the digHoloObject |
|---|---|---|
| in | *cal* | : Pointer to a polCount x batchCount array of complex numbers to be multiplied to each reconstructed field within a processing batch. |
| in | *polCount* | : The number of polarisation components in the supplied calibration. To specify a polarisation-independent calibration, set polCount=1. |
| in | *batchCount* | : The number of batch elements within the calibration. |

**Returns**

errorCode : [DIGHOLO_ERROR_SUCCESS, DIGHOLO_ERROR_INVALIDHANDLE DIGHOLO_ERROR_NULLPOINTER DIGHOLO_ERROR_INVALIDARGUMENT]

### int digHoloConfigSetBatchCalibrationEnabled (int *handleIdx*, int *enabled*)

Sets whether the batchCalibration is enabled or disabled.

See digHoloConfigSetBatchCalibration for further information.

**Parameters**

| in | *handleIdx* | : enumerated handle index of the digHoloObject |
|---|---|---|
| in | *enabled* | : Enable (1) or disable (0) the batchCalibration if available. |

**Returns**

errorCode : [DIGHOLO_ERROR_SUCCESS DIGHOLO_ERROR_INVALIDHANDLE]

### int digHoloConfigSetBatchCalibrationFromFile (int *handleIdx*, const char * *fname*, int *polCount*, int *batchCount*)

Defines an array polCount x batchCount of complex numbers that will be multiplied to each reconstructed field.

See [digHoloConfigSetBatchCalibration()](). This routine is essentially the same, except rather than specifying a pointer as the source of the calibration array, a path to a binary file is provided.

**Parameters**

| in | *handleIdx* | : enumerated handle index of the digHoloObject |
|---|---|---|
| in | *fname* | : Path to a binary file containing the batchCalibration array |
| in | *polCount* | : The number of polarisation components in the supplied calibration. To specify a polarisation-independent calibration, set polCount=1. |
| in | *batchCount* | : The number of batch elements within the calibration. |

**Returns**

errorCode : [DIGHOLO_ERROR_SUCCESS, DIGHOLO_ERROR_INVALIDHANDLE DIGHOLO_ERROR_FILENOTFOUND DIGHOLO_ERROR_INVALIDARGUMENT]

### int digHoloConfigSetBatchCount (int *handleIdx*, int *batchCount*)

Sets the current batch size for processing frames.

Also see digHoloSetBatch().

**Parameters**

| in | *handleIdx* | : enumerated handle index of the digHoloObject |
|---|---|---|
| in | *batchCount* | : the number of frames to be processed. batchCount<1 will default to 1, but will not raise an error. |

**Returns**

    errorCode : [DIGHOLO_ERROR_SUCCESS, DIGHOLO_ERROR_INVALIDHANDLE]

### int digHoloSetBatch (int *handleIdx*, int *batchCount*, float * *frameBuffer*)

Sets the parameters for processing a batch of batchCount frames stored in a float32 frameBuffer.

This function, or a similar 'digHoloSetBatch*()' function should be called before calling digHoloProcessBatch()

**Parameters**

| in | *handleIdx* | : enumerated handle index of the digHoloObject |
|---|---|---|
| in | *batchCount* | : the number of frames to be processed. batchCount<1 will default to 1, but will not raise an error. |
| in | *frameBuffer* | : a pointer to a buffer of frames in float32 format. |

**Returns**

    errorCode : [DIGHOLO_ERROR_SUCCESS, DIGHOLO_ERROR_INVALIDHANDLE, DIGHOLO_ERROR_NULLPOINTER].

### int digHoloSetBatchAvg (int *handleIdx*, int *batchCount*, float * *frameBuffer*, int *avgCount*, int *avgMode*)

Sets the parameters for processing a batch of (batchCount x avgCount) frames stored in a float32 frameBuffer, with averaging.

This function, or a similar 'digHoloSetBatch*()' function should be called before calling digHoloProcessBatch()

Processing can be performed to average multiple measured frames into a single output reconstructed field. To do so, (batchCount x avgCount) FFTs are performed. The off-axis terms in the Fourier plane within the Fourier window are then added constructively (in-phase with one another) within an averaging group, and batchCount IFFTs are performed. Fields are added constructively under the assumption that the reference beam's phase made drift during measurement, and hence must be compensated for by applying phase-shifts to ensure fields within an averaging group are in-phase. There are two averaging modes. 0 : Averaging over blocks of sequential frames (AAABBBCCCDDD). 1: Interlaced frames (ABCDABCDABCD). 2: Averaging over blocks of sequential frequency sweeps. ({A}{A}{A}{B}{B}{B}{C}{C}{C}), where {A} is a wavelength sweep. If wavelength count is 1, this mode is the same as 0.

**Parameters**

| in | *handleIdx* | : enumerated handle index of the digHoloObject |
|---|---|---|
| in | *batchCount* | : the number of unique frames to be processed. batchCount<1 will default to 1, but will not raise an error. |
| in | *frameBuffer* | : a pointer to a buffer of frames in float32 format. |
| in | *avgCount* | : the number of frames which should be averaged together. |
| in | *avgMode* | : Average frames in sequential blocks of avgCount frames (0), or interlaced sequences of avgCount frames (1), sequential blocks of wavelength sweeps of avgCount frams (2) |

**Returns**

errorCode : [DIGHOLO_ERROR_SUCCESS, DIGHOLO_ERROR_INVALIDHANDLE, DIGHOLO_ERROR_NULLPOINTER].

### int digHoloSetBatchAvgUint16 (int *handleIdx*, int *batchCount*, unsigned short * *frameBuffer*, int *avgCount*, int *avgMode*, int *transpose*)

Sets the parameters for processing a batch of batchCount frames stored in a uint16 frameBuffer.

This function, or a similar 'digHoloSetBatch*()' function should be called before calling digHoloProcessBatch()

The equivalent float32 routine (digHoloSetBatchAvg) is the preferred method of inputting frame data as there is a ~20%? performance penalty due to the required conversion from uint16->float32. As opposed to float32 data, which can be fed directly into the FFT, uint16 data must be converted first, and as the buffer is external, the digHoloObject has no way of knowing if the data has changed since the last conversion, and hence must reconvert. Unlike float32. Calling a float32 digHoloSetBatch* routine or digHoloSetFrameBuffer() will reset the frame source as a float32 buffer.

See digHoloSetBatchAvg() for further details on averaging.

**Parameters**

| in | *handleIdx* | : enumerated handle index of the digHoloObject |
|---|---|---|
| in | *batchCount* | : the number of frames to be processed. batchCount<1 will default to 1, but will not raise an error. |
| in | *frameBuffer* | : a pointer to a buffer of frames in uint16 format. |
| in | *avgCount* | : the number of frames which should be averaged together. |
| in | *avgMode* | : Average frames in sequential blocks of avgCount frames (0), or interlaced sequences of avgCount frames (1). |
| in | *transpose* | : setting this flag to true will transpose the individual camera frames during the uint16->float32 format. |

**Returns**

errorCode : [DIGHOLO_ERROR_SUCCESS, DIGHOLO_ERROR_INVALIDHANDLE, DIGHOLO_ERROR_NULLPOINTER].

### int digHoloSetBatchUint16 (int *handleIdx*, int *batchCount*, unsigned short * *frameBuffer*, int *transpose*)

Sets the parameters for processing a batch of batchCount frames stored in a uint16 frameBuffer.

This function, or a similar 'digHoloSetBatch*()' function should be called before calling digHoloProcessBatch()

The equivalent float32 routine (digHoloSetBatch) is the preferred method of inputting frame data as there is a ~20%? performance penalty due to the required conversion from uint16->float32. As opposed to float32 data, which can be fed directly into the FFT, uint16 data must be converted first, and as the buffer is external, the digHoloObject has no way of knowing if the data has changed since the last conversion, and hence must reconvert. Unlike float32. Calling a float32 digHoloSetBatch* routine or digHoloSetFrameBuffer() will reset the frame source as a float32 buffer.

**Parameters**

| in | *handleIdx* | : enumerated handle index of the digHoloObject |
|---|---|---|
| in | *batchCount* | : the number of frames to be processed. batchCount<1 will default to 1, but will not raise an error. |
| in | *frameBuffer* | : a pointer to a buffer of frames in uint16 format. |
| in | *transpose* | : setting this flag to true will transpose the individual camera frames |

| | | during the uint16->float32 format. |
|---|---|---|

**Returns**

    errorCode : [DIGHOLO_ERROR_SUCCESS, DIGHOLO_ERROR_INVALIDHANDLE, DIGHOLO_ERROR_NULLPOINTER].

## Averaging modes

The reconstructed fields can have averaging applied, whereby multiple frames are constructively averaged to produce a reconstructed field with less noise.

### Macros

- #define DIGHOLO_AVGMODE_SEQUENTIAL   0
- #define DIGHOLO_AVGMODE_INTERLACED   1
- #define DIGHOLO_AVGMODE_SEQUENTIALSWEEP   2
- #define DIGHOLO_AVGMODE_COUNT   3

### Detailed Description

The reconstructed fields can have averaging applied, whereby multiple frames are constructively averaged to produce a reconstructed field with less noise.

These defines how the frames to be averaged are arranged in memory.

### Macro Definition Documentation

#### #define DIGHOLO_AVGMODE_COUNT   3

The total number of different averaging modes supported.

#### #define DIGHOLO_AVGMODE_INTERLACED   1

Interlaced averaging. Batches to be averaged are adjacent in memory. Frames to be averaged are batchCount apart. [A B C A B C A B C]

#### #define DIGHOLO_AVGMODE_SEQUENTIAL   0

Sequential averaging. Frames to be averaged are adjacent in memory [A A A B B B C C C]

#### #define DIGHOLO_AVGMODE_SEQUENTIALSWEEP   2

Sequential wavelength sweep averaging. Batches to be averaged are adjacent in memory. Frames to be averaged are wavelengthCount apart [{A} {A} {A} {B} {B} {B} {C} {C} {C}]. Where {A} is a full wavelength sweep. If wavelength count is 1, this is the same as AVGMODE_SEQUENTIAL

# Window sizes

Configuring the size of the windows in the frame plane and Fourier plane, as well as wavelength of operation.

## Functions

- int digHoloConfigSetfftWindowSize (int handleIdx, int width, int height)
- int digHoloConfigGetfftWindowSize (int handleIdx, int *width, int *height)
- int digHoloConfigSetfftWindowSizeX (int handleIdx, int width)
- int digHoloConfigGetfftWindowSizeX (int handleIdx)
- int digHoloConfigGetfftWindowSizeY (int handleIdx)
- int digHoloConfigSetfftWindowSizeY (int handleIdx, int height)
- int digHoloConfigSetFourierWindowRadius (int handleIdx, float windowRadius)
- float digHoloConfigGetFourierWindowRadius (int handleIdx)
- int digHoloConfigSetIFFTResolutionMode (int handleIdx, int resolutionModeIdx)
- int digHoloConfigGetIFFTResolutionMode (int handleIdx)

## Detailed Description

Configuring the size of the windows in the frame plane and Fourier plane, as well as wavelength of operation.

## Function Documentation

### int digHoloConfigGetfftWindowSize (int *handleIdx*, int * *width*, int * *height*)

Gets the width (x-axis) and height (y-axis) of the window within the full-frame which will be FFT'd. Will be multiple of 16 (DIGHOLO_PIXEL_QUANTA).

For example, the full camera frames may be 640x512, but there is only a 256x256 region of interest (window) within that full-frame.

#### Parameters

| | | |
|---|---|---|
| in | *handleIdx* | : enumerated handle index of the digHoloObject |
| out | *width* | : width (x-axis) of the FFT window. |
| out | *height* | : height (y-axis) of the FFT window. |

#### Returns

errorCode : [DIGHOLO_ERROR_SUCCESS, DIGHOLO_ERROR_INVALIDHANDLE]

### int digHoloConfigGetfftWindowSizeX (int *handleIdx*)

Returns the width (x-axis) of the window within the full-frame which will be FFT'd.

#### Parameters

| | | |
|---|---|---|
| in | *handleIdx* | : enumerated handle index of the digHoloObject |

#### Returns

width of FFT window, or zero if handle index is invalid

### int digHoloConfigGetfftWindowSizeY (int *handleIdx*)

Sets the height (y-axis) of the window within the full-frame which will be FFT'd. Must be multiple of 16 (DIGHOLO_PIXEL_QUANTA).

For example, the full camera frames may be 640x512, but there is only a 256x256 region of interest (window) within that full-frame.

**Parameters**

| | | |
|---|---|---|
| in | *handleIdx* | : enumerated handle index of the digHoloObject |

**Returns**

    height of FFT window, or zero if handle index is invalid

### float digHoloConfigGetFourierWindowRadius (int *handleIdx*)

Returns the current Fourier window radius in degrees.

**Parameters**

| | | |
|---|---|---|
| in | *handleIdx* | : enumerated handle index of the digHoloObject |

**Returns**

    radius of Fourier window (or zero if invalid handleIdx)

### int digHoloConfigGetIFFTResolutionMode (int *handleIdx*)

Gets whether the IFFT has the same dimensions as the FFT (0), or of the smaller Fourier window (1)

0 : IFFT has the same pixel number and dimensions as the FFT (fftWindowSizeX/Y), and hence the reconstructed field is the same dimensions as the original camera frame. This mode yields no extra information and is slower, but it is sometimes conveinient. 1 : Recommended mode. IFFT has the minimum dimensions, those of the Fourier window. Faster with no loss of information.

**Parameters**

| | | |
|---|---|---|
| in | *handleIdx* | : enumerated handle index of the digHoloObject |

**Returns**

    resolutionModeIdx : Same as FFT dimensions (0), Fourier window dimensions (1). Returns zero for invalid handle index

### int digHoloConfigSetfftWindowSize (int *handleIdx*, int *width*, int *height*)

Sets the width (x-axis) and height (y-axis) of the window within the full-frame which will be FFT'd. Must be multiple of 16 (DIGHOLO_PIXEL_QUANTA).

For example, the full camera frames may be 640x512, but there is only a 256x256 region of interest (window) within that full-frame.

**Parameters**

| | | |
|---|---|---|
| in | *handleIdx* | : enumerated handle index of the digHoloObject |
| in | *width* | : width (x-axis) of the FFT window. |
| in | *height* | : height (y-axis) of the FFT window. |

**Returns**

    errorCode : [DIGHOLO_ERROR_SUCCESS, DIGHOLO_ERROR_INVALIDHANDLE, DIGHOLO_ERROR_INVALIDDIMENSION]

### int digHoloConfigSetfftWindowSizeX (int *handleIdx*, int *width*)

Sets the width (x-axis) of the window within the full-frame which will be FFT'd. Must be multiple of 16 (DIGHOLO_PIXEL_QUANTA).

For example, the full camera frames may be 640x512, but there is only a 256x256 region of interest (window) within that full-frame.

**Parameters**

| | | |
|---|---|---|
| in | *handleIdx* | : enumerated handle index of the digHoloObject |
| in | *width* | : width (x-axis) of the FFT window. |

**Returns**

    width : The width actually set, or zero if error. Values will be floored to nearest multiple of 16 (DIGHOLO_PIXEL_QUANTA), or will return 0 if invalid handle or negative/zero size is requested.

### int digHoloConfigSetfftWindowSizeY (int *handleIdx*, int *height*)

Returns the height (y-axis) of the window within the full-frame which will be FFT'd.

**Parameters**

| | | |
|---|---|---|
| in | *handleIdx* | : enumerated handle index of the digHoloObject |
| in | *height* | height (y-axis) of the FFT window. |

**Returns**

    height : The height actually set, or zero if error. Values will be floored to nearest multiple of 16 (DIGHOLO_PIXEL_QUANTA), or will return 0 if invalid handle or negative/zero size is requested.

### int digHoloConfigSetFourierWindowRadius (int *handleIdx*, float *windowRadius*)

Sets the radius (in degrees) of the circular window in Fourier space which selects the off-axis term, that will then be IFFT'd to reconstruct the field.

This is the circular window in Fourier space that is selected. e.g. w_c https://doi.org/10.1364/OE.25.033400 Fig. 1 The radius would typically be no more than ~1/3 [sqrt(2)/(3+sqrt(2))]=0.320513 of the maximum resolvable angle set by the frame pixel size. (e.g. Fig 1c) e.g. if the wavelength is 1565e-9 and the pixelSize is 20e-6, w_max would be (1565e-9/(2*20e-6))*(180/pi) = 2.24 degrees, and window radius (w_c) should be less than 0.3205*2.24=0.719 degrees The reference beam tilt angle in x and y would have to be (x,y)=(3w_c,3w_c)/sqrt(2)=(1.525,1.525) degrees. If the full resolution of the camera is not required, smaller windows and shallower reference angles can be employed. Smaller windows are also less likely to capture unwanted noise. If Fourier wrapping (e.g. Fig 1b) is employed, a larger fractional window could be used [+ root of 0=8w^2+2-2]->[(-2+sqrt(68))/16]=0.3904. Tilt (x,y) = (w_max-w_c,w_max) Formerly called 'digHoloSetApertureSize'

**Parameters**

| | | |
|---|---|---|
| in | *handleIdx* | : enumerated handle index of the digHoloObject |
| in | *windowRadius* | : Radius of the Fourier window in degrees. |

**Returns**

    errorCode : [DIGHOLO_ERROR_SUCCESS, DIGHOLO_ERROR_INVALIDHANDLE]

### int digHoloConfigSetIFFTResolutionMode (int *handleIdx*, int *resolutionModeIdx*)

Sets whether the IFFT has the same dimensions as the FFT (0), or of the smaller Fourier window (1)

0 : IFFT has the same pixel number and dimensions as the FFT (fftWindowSizeX/Y), and hence the reconstructed field is the same dimensions as the original camera frame. This mode yields no extra information and is slower, but it is sometimes conveinient. 1 : Recommended mode. IFFT has the minimum dimensions, those of the Fourier window. Faster with no loss of information.

**Parameters**

| | | |
|---|---|---|
| in | *handleIdx* | : enumerated handle index of the digHoloObject |
| in | *resolutionModeIdx* | : Same as FFT dimensions (0), Fourier window dimensions (1). |

**Returns**

errorCode : [DIGHOLO_ERROR_SUCCESS, DIGHOLO_ERROR_INVALIDHANDLE, DIGHOLO_ERROR_INVALIDARGUMENT]

# Wavelength

Configuring the current operating wavelength. (e.g. for converting angles to k-space)

## Modules

- Input/output wavelength ordering
  *Specifies how data corresponding to wavelength sweeps are organised in memory.*

## Functions

- int digHoloConfigSetWavelengthCentre (int handleIdx, float lambda0)
- float digHoloConfigGetWavelengthCentre (int handleIdx)
- int digHoloConfigSetWavelengths (int handleIdx, float *wavelengths, int wavelengthCount)
- int digHoloConfigSetWavelengthsLinearFrequency (int handleIdx, float wavelengthStart, float wavelengthStop, int wavelengthCount)
- float * digHoloConfigGetWavelengths (int handleIdx, int *wavelengthCount)
- int digHoloConfigSetWavelengthOrdering (int handleIdx, int inoutIdx, int ordering)
- int digHoloConfigGetWavelengthOrdering (int handleIdx, int inoutIdx)

## Detailed Description

Configuring the current operating wavelength. (e.g. for converting angles to k-space)

## Function Documentation

### float digHoloConfigGetWavelengthCentre (int *handleIdx*)

Gets the default centre wavelength. The dimensions in Fourier space in terms of angles will depend on the operating wavelength.

**Parameters**

| in | *handleIdx* | : enumerated handle index of the digHoloObject |
|---|---|---|

**Returns**

lambda0 : Operating wavelength. Returns zero for invalid handle index.

### int digHoloConfigGetWavelengthOrdering (int *handleIdx*, int *inoutIdx*)

Sets the ordering of wavelength data (Fast vs. slow) for the input frame source, or the output fields and coefficients.

**Parameters**

| in | *handleIdx* | : enumerated handle index of the digHoloObject |
|---|---|---|
|  | *inoutIdx* | : Values apply to the input source (DIGHOLO_WAVELENGTHORDER_INPUT) or the output fields and coefficients (DIGHOLO_WAVELENGTHORDER_OUTPUT) |

**Returns**

ordering : Wavelength is the fast changing axis (DIGHOLO_WAVELENGTHORDER_FAST) or the slow changing axis (DIGHOLO_WAVELENGTHORDER_SLOW). Invalid handle or invalid input/output specification will return 0.

**float \* digHoloConfigGetWavelengths (int   handleIdx, int \*   wavelengthCount)**

Returns an array corresponding to the current wavelength axis.

Note, this may only be updated after digital holography processing has been performed.

**Parameters**

| in | *handleIdx* | : enumerated handle index of the digHoloObject |
|---|---|---|
|  | *wavelengthCount* | : Pointer to an int32 where the length of the 'wavelengths' array will be returned. |

**Returns**

wavelengths : Pointer to an array containing wavelengthCount wavelengths

**int digHoloConfigSetWavelengthCentre (int   handleIdx, float   lambda0)**

Sets the default centre wavelength. The dimensions in Fourier space in terms of angles will depend on the operating wavelength.

**Parameters**

| in | *handleIdx* | : enumerated handle index of the digHoloObject |
|---|---|---|
|  | *lambda0* | : Operating wavelength. |

**Returns**

errorCode : [DIGHOLORERROR_SUCCESS, DIGHOLO_ERROR_INVALIDHANDLE, DIGHOLO_ERROR_INVALIDARGUMENT]

**int digHoloConfigSetWavelengthOrdering (int   handleIdx, int   inoutIdx, int   ordering)**

Sets the ordering of wavelength data (Fast vs. slow) for the input frame source, or the output fields and coefficients.

**Parameters**

| in | *handleIdx* | : enumerated handle index of the digHoloObject |
|---|---|---|
|  | *inoutIdx* | : Values apply to the input source (DIGHOLO_WAVELENGTHORDER_INPUT) or the output fields and coefficients (DIGHOLO_WAVELENGTHORDER_OUTPUT) |
|  | *ordering* | : Wavelength is the fast changing axis (DIGHOLO_WAVELENGTHORDER_FAST) or the slow changing axis (DIGHOLO_WAVELENGTHORDER_SLOW) |

**Returns**

errorCode : [DIGHOLORERROR_SUCCESS, DIGHOLO_ERROR_INVALIDHANDLE, DIGHOLO_ERROR_INVALIDDARGUMENT]

**int digHoloConfigSetWavelengths (int   handleIdx, float \*   wavelengths, int   wavelengthCount)**

Sets the wavelength axis for wavelength-dependent batch processing. This is for use when the frames captured are at different wavelengths.

**Parameters**

| in | *handleIdx* | : enumerated handle index of the digHoloObject |
|---|---|---|
|  | *wavelengths* | : Pointer to an array containing wavelengthCount wavelengths |
|  | *wavelengthCount* | : The length of the wavelengths array |

**Returns**

errorCode : [DIGHOLORERROR_SUCCESS, DIGHOLO_ERROR_INVALIDHANDLE, DIGHOLO_ERROR_INVALIDDIMENSION, DIGHOLO_ERROR_NULLPOINTER]

**int digHoloConfigSetWavelengthsLinearFrequency (int** *handleIdx***, float** *wavelengthStart***, float** *wavelengthStop***, int** *wavelengthCount***)**

Sets a wavelength axis that is linear in frequency between wavelengthStart and wavelengthStop of length wavelengthCount

**Parameters**

| in | *handleIdx* | : enumerated handle index of the digHoloObject |
|---|---|---|
| | *wavelengthStart* | : The starting wavelength of the linear frequency sweep |
| | *wavelengthStop* | : The final wavelength of the linear frequency sweep |
| | *wavelengthCount* | : The total number of wavelengths |

**Returns**

errorCode : [DIGHOLORERROR_SUCCESS, DIGHOLO_ERROR_INVALIDHANDLE, DIGHOLO_ERROR_INVALIDDIMENSION]

# Input/output wavelength ordering

Specifies how data corresponding to wavelength sweeps are organised in memory.

## Macros

- #define DIGHOLO_WAVELENGTHORDER_INPUT   0
- #define DIGHOLO_WAVELENGTHORDER_OUTPUT   1
- #define DIGHOLO_WAVELENGTHORDER_FAST   0
- #define DIGHOLO_WAVELENGTHORDER_SLOW   1

## Detailed Description

Specifies how data corresponding to wavelength sweeps are organised in memory.

The user can input frames where adjacent frames in memory are at different wavelengths, or the same wavelength. The user can also specify how to arrange the fields and output mode basis coefficients.

For example, the user may want to experimentally capture frames where an input is kept constant, and the wavelength of the source is swept (Input = ORDER_FAST), however the output reconstructed fields may be desired such that transfer matrices for each wavelength are output in continous blocks of memory (Output = ORDER_SLOW).

Named 'fast' and 'slow' because the wavelength axis is either the fast changing axis, or the slowly changing axis in the data.

## Macro Definition Documentation

### #define DIGHOLO_WAVELENGTHORDER_FAST   0

Wavelength is the fast changing axis. Data corresponding to different wavelengths is adjacent in memory and changes from one element to the next.

### #define DIGHOLO_WAVELENGTHORDER_INPUT   0

Input. The ordering of the input frames.

### #define DIGHOLO_WAVELENGTHORDER_OUTPUT   1

Output. The ordering of the output fields and modal coefficients.

### #define DIGHOLO_WAVELENGTHORDER_SLOW   1

Wavelength is the slow changing axis. Data corresponding to the same wavelength is adjacent in memory. All data for a particular wavelength is continous in memory.

# Alignment

Configuring key-properties such as reference beam tilt, defocus, beam centre, etc.)

## Modules

- Automatic (AutoAlign)
  *Automatic search for the optimal alignment settings for current batch. (e.g. tilt, beam centre etc)*
- Set/Get parameters
  *Set or retrieve current alignment settings.*

## Detailed Description

Configuring key-properties such as reference beam tilt, defocus, beam centre, etc.)

# Automatic (AutoAlign)

Automatic search for the optimal alignment settings for current batch. (e.g. tilt, beam centre etc)

## Modules

- [AutoAlign operation modes](#)
  *Specifies which type of AutoAlign routine to perform.*

## Functions

- int [digHoloConfigSetAutoAlignBeamCentre](#) (int handleIdx, int enable)
  *Sets whether beam centering optimisation in the camera plane should be performed by the AutoAlign routine.*

- int [digHoloConfigGetAutoAlignBeamCentre](#) (int handleIdx)
  *Returns whether beam centering optimisation is enabled/disabled during the AutoAlign routine.*

- int [digHoloConfigSetAutoAlignDefocus](#) (int handleIdx, int enable)
  *Sets whether defocus optimisation in the camera plane should be performed by the AutoAlign routine.*

- int [digHoloConfigGetAutoAlignDefocus](#) (int handleIdx)
  *Returns whether defocus optimisation is enabled/disabled during the AutoAlign routine.*

- int [digHoloConfigSetAutoAlignTilt](#) (int handleIdx, int enable)
  *Sets whether tilt optimisation in the camera plane (off-axis position in the Fourier plane) should be performed by the AutoAlign routine.*

- int [digHoloConfigGetAutoAlignTilt](#) (int handleIdx)
  *Returns whether tilt optimisation is enabled/disabled during the AutoAlign routine.*

- int [digHoloConfigSetAutoAlignBasisWaist](#) (int handleIdx, int enable)
  *Sets whether basis waist optimisation in the camera plane should be performed by the AutoAlign routine.*

- int [digHoloConfigGetAutoAlignBasisWaist](#) (int handleIdx)
  *Returns whether basis waist optimisation is enabled/disabled during the AutoAlign routine.*

- int [digHoloConfigSetAutoAlignFourierWindowRadius](#) (int handleIdx, int enable)
  *Sets whether Fourier window radius optimisation in the camera plane should be performed by the AutoAlign routine.*

- int [digHoloConfigGetAutoAlignFourierWindowRadius](#) (int handleIdx)
  *Returns whether Fourier window radius optimisation is enabled/disabled during the AutoAlign routine.*

- int [digHoloConfigSetAutoAlignTol](#) (int handleIdx, float tolerance)
  *Sets the convergence tolerance for the AutoAlign routine.*

- float digHoloConfigGetAutoAlignTol (int handleIdx)
  *Returns the convergence tolerance for the AutoAlign routine.*

- int digHoloConfigSetAutoAlignPolIndependence (int handleIdx, int polIndependence)
  *Defines whether the polarisation components should be treated as combined fields(0), or independent components(1).*

- int digHoloConfigGetAutoAlignPolIndependence (int handleIdx)
  *Returns whether the polarisation components are treated as combined fields(0), or independent components(1).*

- int digHoloConfigSetAutoAlignBasisMulConjTrans (int handleIdx, int mulConjTranspose)
  *Defines whether properties of the transfer matrix such as 'diagonal' and 'crosstalk' are calculated for the raw transfer matrix (A) or for (AA*)*

- int digHoloConfigGetAutoAlignBasisMulConjTrans (int handleIdx)
  *Returns whether properties of the transfer matrix such as 'diagonal' and 'crosstalk' are calculated for the raw transfer matrix (0=A) or for (1=AA*)*

- int digHoloConfigSetAutoAlignMode (int handleIdx, int alignMode)
  *Sets what type of AutoAlignment routine to perform when digHoloAutoAlign() is run.*

- int digHoloConfigGetAutoAlignMode (int handleIdx)
  *Returns the current AutoAlign mode [Full, Tweak, Estimate].*

- int digHoloConfigSetAutoAlignGoalIdx (int handleIdx, int goalIdx)
  *Specifies which batch metric to use for optimisation during the AutoAlign routine.*

- int digHoloConfigGetAutoAlignGoalIdx (int handleIdx)
  *Returns the current metric used for optimisation during the AutoAlignRoutine.*

- float digHoloAutoAlign (int handleIdx)
  *Runs the AutoAlign routine that searches for optimal settings [e.g. tilt, defocus, beam centre, waist] with respect to the specfied goalIdx [e.g. IL, MDL, SNR etc].*

- float digHoloAutoAlignGetMetric (int handleIdx, int metricIdx)
  *Returns the specified metric, if available, from the last AutoAlign routine run.*

- float * digHoloAutoAlignGetMetrics (int handleIdx, int metricIdx)
  *Returns a pointer to an array containing metrics, if available, from the last AutoAlign routine run. The preferred method is the safer digHoloAutoAlignGetMetric().*

- int digHoloAutoAlignCalcMetrics (int handleIdx)
  *Calculates the AutoAlign metrics, from the most recent available coefficients.*

## Detailed Description

Automatic search for the optimal alignment settings for current batch. (e.g. tilt, beam centre etc)

## Function Documentation

### float digHoloAutoAlign (int *handleIdx*)

Runs the AutoAlign routine that searches for optimal settings [e.g. tilt, defocus, beam centre, waist] with respect to the specfied goalIdx [e.g. IL, MDL, SNR etc].

The AutoAlign routine has two parts. A first section which attempts to rapidly estimate the parameters (tilt,defocus,beam centre, waist) from scratch, followed by a fine alignment tweaking stage. The alignment mode specifies whether to run both stages, or just 1 of the stages of autoalignment.

Alignment routine assumes that polarisation components are mostly on opposite sides of the frame along the x-axis (width), which is typically the widest axis of a camera.

See digHoloConfigSetAutoAlignMode() for types of AutoAlignment routines. See digHoloConfigSetAutoAlignGoalIdx() for types of AutoAlignment goal parameters for optimisation.

**Parameters**

| in | *handleIdx* | : enumerated handle index of the digHoloObject |
|---|---|---|

**Returns**

metricValue : The final value for the specified goal metric. Other metrics that may also be available can be seen using the digHolo

### int digHoloAutoAlignCalcMetrics (int *handleIdx*)

Calculates the AutoAlign metrics, from the most recent available coefficients.

This function can be used to update the AutoAlignMetrics after performing Batch processing (e.g. digHoloProcessBatch()). Calling this function will analyse the current coeffients and extract the same metrics the AutoAlign routine uses, without having to run a full AutoAlign. Most likely use case is updating the coefficients using digHoloProcessBatch, and then calling digHoloAutoAlignGetMetric() or digHoloAutoAlignGetMetrics() to get the corresponding metrics for the current batch.

**Parameters**

| in | *handleIdx* | : enumerated handle index of the digHoloObject |
|---|---|---|

**Returns**

errorCode : [DIGHOLO_ERROR_SUCCESS, DIGHOLO_ERROR_INVALIDHANDLE, DIGHOLOERROR_NULLPOINTER]. Function will return NULLPOINTER if no coefficients exist because AutoAlign nor ProcessBatch type routines have every been run previously.

### float digHoloAutoAlignGetMetric (int *handleIdx*, int *metricIdx*)

Returns the specified metric, if available, from the last AutoAlign routine run.

digHoloAutoAlign() optimises with respect to a specific goal metric, however it will also calculate the other metrics at the end. This allows the user to access those metrics. If multiple wavelengths are involved, this will return the average value over all wavelengths.

**Parameters**

| | | |
|---|---|---|
| in | *handleIdx* | : enumerated handle index of the digHoloObject |
| in | *metricIdx* | : DIGHOLO_METRIC_[IL, MDL, DIAG, SNRAVG, DIAGBEST, DIAGWORST, SNRBEST, SNRWORST, SNRMG]. |

**Returns**

metricValue : The metric as calculated from the most recent digHoloAutoAlign() run. Will return 0 if an invalid handle index or invalid metric index is requested. A value of -FLT_MAX (~-3.4e38) means the metric has not been calculated.

## float * digHoloAutoAlignGetMetrics (int *handleIdx*, int *metricIdx*)

Returns a pointer to an array containing metrics, if available, from the last AutoAlign routine run. The preferred method is the safer digHoloAutoAlignGetMetric().

digHoloAutoAlign() optimises with respect to a specific goal metric, however it will also calculate the other metrics at the end. This allows the user to access those metrics.

**Parameters**

| | | |
|---|---|---|
| in | *handleIdx* | : enumerated handle index of the digHoloObject |
| in | *metricIdx* | : DIGHOLO_METRIC_[IL, MDL, DIAG, SNRAVG, DIAGBEST, DIAGWORST, SNRBEST, SNRWORST, SNRMG]. |

**Returns**

metricValue[] : Array of all metrics as calculated from the most recent digHoloAutoAlign() run. This array will be of digHoloWavelengthCount+1 long. Where the first digHoloWavelengthCount elements are the values calculated at each wavelength, and the final [digHoloWavelengthCount] element is the average over all wavelengths. Will return nullptr if an invalid handle index. A value of -FLT_MAX (~-3.4e38) means the metric has not been calculated.

## int digHoloConfigGetAutoAlignBasisMulConjTrans (int *handleIdx*)

Returns whether properties of the transfer matrix such as 'diagonal' and 'crosstalk' are calculated for the raw transfer matrix (0=A) or for (1=AA*)

Parameters such as the diagonal and the crosstalk of a transfer matrix for a given batch, will depend on the spatial basis. However in practice, it may not always be conveinient to measure in that spatial basis, or the correct spatial basis that approximately diagonalises the transfer matrix may not be known. In such a scenario, it can be conveinient to calculate parameters like the diagonal and crosstalk using the result of the matrix multiplied by it's conjugate transpose. For an approximately unitary transfer matrix, this will yield an approximately diagonal matrix, from which ~diagonal and ~crosstalk like values can be extracted, without having to measure directly in the correct spatial basis.

**Parameters**

| | | |
|---|---|---|
| in | *handleIdx* | : enumerated handle index of the digHoloObject return mulConjTranspose : Parameters calculated for raw matrix [A] (0). Calculated for AA* (1). Will return zero for invalid handle index. |

## int digHoloConfigGetAutoAlignBasisWaist (int *handleIdx*)

Returns whether basis waist optimisation is enabled/disabled during the AutoAlign routine.

**Parameters**

| in | *handleIdx* | : enumerated handle index of the digHoloObject |
|---|---|---|

**Returns**

enabled : [true/false], will return false if handle index is invalid.

### int digHoloConfigGetAutoAlignBeamCentre (int  *handleIdx*)

Returns whether beam centering optimisation is enabled/disabled during the AutoAlign routine.

**Parameters**

| in | *handleIdx* | : enumerated handle index of the digHoloObject |
|---|---|---|

**Returns**

enabled : [true/false], will return false if handle index is invalid.

### int digHoloConfigGetAutoAlignDefocus (int  *handleIdx*)

Returns whether defocus optimisation is enabled/disabled during the AutoAlign routine.

**Parameters**

| in | *handleIdx* | : enumerated handle index of the digHoloObject |
|---|---|---|

**Returns**

enabled : [true/false], will return false if handle index is invalid.

### int digHoloConfigGetAutoAlignFourierWindowRadius (int  *handleIdx*)

Returns whether Fourier window radius optimisation is enabled/disabled during the AutoAlign routine.

**Parameters**

| in | *handleIdx* | : enumerated handle index of the digHoloObject |
|---|---|---|

**Returns**

enabled : [true/false], will return false if handle index is invalid.

### int digHoloConfigGetAutoAlignGoalIdx (int  *handleIdx*)

Returns the current metric used for optimisation during the AutoAlignRoutine.

**Parameters**

| in | *handleIdx* | : enumerated handle index of the digHoloObject |
|---|---|---|

**Returns**

goalIdx : DIGHOLO_METRIC_[IL, MDL, DIAG, SNRAVG, DIAGBEST, DIAGWORST, SNRBEST, SNRWORST, SNRMG]. Will return 0 if handle index is invalid

### int digHoloConfigGetAutoAlignMode (int *handleIdx*)

Returns the current AutoAlign mode [Full, Tweak, Estimate].

**Parameters**

| in | *handleIdx* | : enumerated handle index of the digHoloObject |
|---|---|---|

**Returns**
  alignMode : Current AutoAlign mode. Will return 0 if handle index is invalid.

### int digHoloConfigGetAutoAlignPolIndependence (int *handleIdx*)

Returns whether the polarisation components are treated as combined fields(0), or independent components(1).

For example, if one polarisation was absent, the mode dependent loss (MDL) would be unaffected for (0) and would be -Inf for (1). For combined fields (0), the transfer matrix is constructed by analysing the batchCount camera frames, and extracting (modeCount x polCount) coefficients per frame. A matrix (batchCount) x (modeCount x polCount) For independent components (1), each polarisation component on the camera is treated as corresponding with a separate input. i.e. a scenario where both polarisation states are launching in simultaneously and can be resolved separately on the camera. In that case, the matrix is (batchCount x polCount) x (modeCount x polCount) with elements in the matrix corresponding with polarisation coupling set to zero. It is as if each polarisation component could have been captured on the camera independently, but was instead captured using a single frame.

**Parameters**

| in | *handleIdx* | : enumerated handle index of the digHoloObject |
|---|---|---|

**Returns**
  polIndependence : Each polarisation component within a frame should be considered part of the same field (0), or a completely independent field (1). Will return zero for invalid handle index.

### int digHoloConfigGetAutoAlignTilt (int *handleIdx*)

Returns whether tilt optimisation is enabled/disabled during the AutoAlign routine.

**Parameters**

| in | *handleIdx* | : enumerated handle index of the digHoloObject |
|---|---|---|

**Returns**
  enabled : [true/false], will return false if handle index is invalid.

### float digHoloConfigGetAutoAlignTol (int *handleIdx*)

Returns the convergence tolerance for the AutoAlign routine.

**Parameters**

| in | *handleIdx* | : enumerated handle index of the digHoloObject |
|---|---|---|

**Returns**

    tolerance : When the metric being optimised fails to improve by this value in one optimisation iteration, the AutoAlign routine will terminate. Returns 0 for if invalid handle index.

**int digHoloConfigSetAutoAlignBasisMulConjTrans (int *handleIdx*, int *mulConjTranspose*)**

Defines whether properties of the transfer matrix such as 'diagonal' and 'crosstalk' are calculated for the raw transfer matrix (A) or for (AA*)

Parameters such as the diagonal and the crosstalk of a transfer matrix for a given batch, will depend on the spatial basis. However in practice, it may not always be conveinient to measure in that spatial basis, or the correct spatial basis that approximately diagonalises the transfer matrix may not be known. In such a scenario, it can be conveinient to calculate parameters like the diagonal and crosstalk using the result of the matrix multiplied by it's conjugate transpose. For an approximately unitary transfer matrix, this will yield an approximately diagonal matrix, from which ~diagonal and ~crosstalk like values can be extracted, without having to measure directly in the correct spatial basis.

**Parameters**

| in | *handleIdx* | : enumerated handle index of the digHoloObject |
|---|---|---|
| in | *mulConjTranspose* | : Parameters calculated for raw matrix [A] (0). Calculated for AA* (1). |

**Returns**

    errorCode : [DIGHOLO_ERROR_SUCCESS, DIGHOLO_ERROR_INVALIDHANDLE]

**int digHoloConfigSetAutoAlignBasisWaist (int *handleIdx*, int *enable*)**

Sets whether basis waist optimisation in the camera plane should be performed by the AutoAlign routine.

**Parameters**

| in | *handleIdx* | : enumerated handle index of the digHoloObject |
|---|---|---|
| in | *enable* | : Enables/disables waist optimisation during the AutoAlign routine. |

**Returns**

    errorCode : [DIGHOLO_ERROR_SUCCESS, DIGHOLO_ERROR_INVALIDHANDLE]

**int digHoloConfigSetAutoAlignBeamCentre (int *handleIdx*, int *enable*)**

Sets whether beam centering optimisation in the camera plane should be performed by the AutoAlign routine.

**Parameters**

| in | *handleIdx* | : enumerated handle index of the digHoloObject |
|---|---|---|
| in | *enable* | : Enables/disables beam centering optimisation during the AutoAlign routine. |

**Returns**

    errorCode : [DIGHOLO_ERROR_SUCCESS, DIGHOLO_ERROR_INVALIDHANDLE]

**int digHoloConfigSetAutoAlignDefocus (int *handleIdx*, int *enable*)**

Sets whether defocus optimisation in the camera plane should be performed by the AutoAlign routine.

**Parameters**

| in | *handleIdx* | : enumerated handle index of the digHoloObject |
|---|---|---|
| in | *enable* | : Enables/disables defocus optimisation during the AutoAlign routine. |

**Returns**

errorCode : [DIGHOLO_ERROR_SUCCESS, DIGHOLO_ERROR_INVALIDHANDLE]

### int digHoloConfigSetAutoAlignFourierWindowRadius (int *handleIdx*, int *enable*)

Sets whether Fourier window radius optimisation in the camera plane should be performed by the AutoAlign routine.

This option simply sets the Fourier window as large as possible for the tilt(x,y).

**Parameters**

| in | *handleIdx* | : enumerated handle index of the digHoloObject |
|---|---|---|
| in | *enable* | : Enables/disables Fourier window radius optimisation during the AutoAlign routine. |

**Returns**

errorCode : [DIGHOLO_ERROR_SUCCESS, DIGHOLO_ERROR_INVALIDHANDLE]

### int digHoloConfigSetAutoAlignGoalIdx (int *handleIdx*, int *goalIdx*)

Specifies which batch metric to use for optimisation during the AutoAlign routine.

DIGHOLO_METRIC_IL 0 //Insertion loss DIGHOLO_METRIC_MDL 1 //Mode dependent loss (a.k.a matrix condition number) DIGHOLO_METRIC_DIAG 2 //the total power along the diagonal of the transfer matrix DIGHOLO_METRIC_SNRAVG 3 //the total power along the diagonal/total power off-diagonal DIGHOLO_METRIC_DIAGBEST 4 //the diagonal element with the most power DIGHOLO_METRIC_DIAGWORST 5 //the diagonal element with the worst power DIGHOLO_METRIC_SNRBEST 6 //the diagonal/off-diagonal power of the best batch element DIGHOLO_METRIC_SNRWORST 7 //the diagonal/off-diagonal power of the worst batch element DIGHOLO_METRIC_SNRMG 8 //Degenerate mode group signal to noise

**Parameters**

| in | *handleIdx* | : enumerated handle index of the digHoloObject |
|---|---|---|
| in | *goalIdx* | : DIGHOLO_METRIC_[IL, MDL, DIAG, SNRAVG, DIAGBEST, DIAGWORST, SNRBEST, SNRWORST, SNRMG] |

**Returns**

errorCode : [DIGHOLO_ERROR_SUCCESS DIGHOLO_ERROR_INVALIDHANDLE DIGHOLO_ERROR_INVALIDARGUMENT]

### int digHoloConfigSetAutoAlignMode (int *handleIdx*, int *alignMode*)

Sets what type of AutoAlignment routine to perform when digHoloAutoAlign() is run.

The AutoAlign routine has two parts. A first section which attempts to rapidly estimate the parameters (tilt,defocus,beam centre, waist) from scratch, followed by a fine alignment tweaking stage. The alignment mode specifies whether to run both stages, or just 1 of the stages of autoalignment.

Alignment routine assumes that polarisation components are mostly on opposite sides of the frame along the x-axis (width), which is typically the widest axis of a camera.

Full (AUTOALIGNMODE_FULL) : Full alignment from scratch, including fine tweaking. Slowest mode, but most accurate. Tweak (AUTOALIGNMODE_TWEAK) : Only runs the second tweak stage, which optimises starting from the current settings. Second slowest mode. Only accurate if the settings are already near the answer. Estimate (AUTOALIGNMODE_ESTIMATE) : Only runs the first 'snap' stage, which rapidly estimates the parameters. Fastest mode. It fairly accurate with tilt, beam centre. If the correct number of mode groups (maxMG) is selected, it is moderately accurate at estimating waist. As currently implemented, does not have the ability to differentiate between +/- defocus, but is moderately accurate for the magnitude.

**Parameters**

| in | handleIdx | : enumerated handle index of the digHoloObject |
|---|---|---|
| in | alignMode | : [AUTOALIGNMODE_FULL,AUTOALIGNMODE_TWEAK,AUTOALIGNMODE_ESTIMATE] |

**Returns**

errorCode : [DIGHOLO_ERROR_SUCCESS DIGHOLO_ERROR_INVALIDHANDLE DIGHOLO_ERROR_INVALIDARGUMENT]

## int digHoloConfigSetAutoAlignPolIndependence (int *handleIdx*, int *polIndependence*)

Defines whether the polarisation components should be treated as combined fields(0), or independent components(1).

For example, if one polarisation was absent, the mode dependent loss (MDL) would be unaffected for (0) and would be -Inf for (1). For combined fields (0), the transfer matrix is constructed by analysing the batchCount camera frames, and extracting (modeCount x polCount) coefficients per frame. A matrix (batchCount) x (modeCount x polCount) For independent components (1), each polarisation component on the camera is treated as corresponding with a separate input. i.e. a scenario where both polarisation states are launching in simultaneously and can be resolved separately on the camera. In that case, the matrix is (batchCount x polCount) x (modeCount x polCount) with elements in the matrix corresponding with polarisation coupling set to zero. It is as if each polarisation component could have been captured on the camera independently, but was instead captured using a single frame.

**Parameters**

| in | handleIdx | : enumerated handle index of the digHoloObject |
|---|---|---|
| in | polIndependence | : Each polarisation component within a frame should be considered part of the same field (0), or a completely independent field (1). |

**Returns**

errorCode : [DIGHOLO_ERROR_SUCCESS, DIGHOLO_ERROR_INVALIDHANDLE]

## int digHoloConfigSetAutoAlignTilt (int *handleIdx*, int *enable*)

Sets whether tilt optimisation in the camera plane (off-axis position in the Fourier plane) should be performed by the AutoAlign routine.

**Parameters**

| in | *handleIdx* | : enumerated handle index of the digHoloObject |
|---|---|---|
| in | *enable* | : Enables/disables tilt optimisation during the AutoAlign routine. |

**Returns**

    errorCode : [DIGHOLO_ERROR_SUCCESS, DIGHOLO_ERROR_INVALIDHANDLE]

### int digHoloConfigSetAutoAlignTol (int   *handleIdx*, float   *tolerance*)

Sets the convergence tolerance for the AutoAlign routine.

**Parameters**

| in | *handleIdx* | : enumerated handle index of the digHoloObject |
|---|---|---|
| in | *tolerance* | : When the metric being optimised fails to improve by this value in one optimisation iteration, the AutoAlign routine will terminate. |

**Returns**

    errorCode : [DIGHOLO_ERROR_SUCCESS, DIGHOLO_ERROR_INVALIDHANDLE]

# AutoAlign operation modes

Specifies which type of AutoAlign routine to perform.

## Macros

- #define DIGHOLO_AUTOALIGNMODE_FULL   0
- #define DIGHOLO_AUTOALIGNMODE_TWEAK   1
- #define DIGHOLO_AUTOALIGNMODE_ESTIMATE   2
- #define DIGHOLO_AUTOALIGNMODE_COUNT   3

## Detailed Description

Specifies which type of AutoAlign routine to perform.

The digHoloAutoAlign routine can operate in 3 ways, set by the digHoloConfigSetAutoAlignMode() function. The 'snap' routine, is routine that can be run at the start of an AutoAlign that discards any existing alignment information (tilt,beam centre, waist, defocus) and takes a best guess at where the off-axis term is in Fourier space. The 'snap' will typically be fairly close to the answer, a finer alignment is then performed for optimal results, at the expense of speed. For optimal performance, try to have minimal defocus on the reference beam.

## Macro Definition Documentation

### #define DIGHOLO_AUTOALIGNMODE_COUNT   3

Total number of different AutoAlign modes

### #define DIGHOLO_AUTOALIGNMODE_ESTIMATE   2

Only do snap routine, don't do any fine-alignment. Useful for speed, or if the beam being analysed is not appropriately described in the HG basis (fine alignment uses HG basis).

### #define DIGHOLO_AUTOALIGNMODE_FULL   0

A full alignment routine is run.

### #define DIGHOLO_AUTOALIGNMODE_TWEAK   1

Snap routine is skipped. This may be desirable if the snap is not locking onto the area of Fourier space you want to reconstruct. When snap is off, the routine will just attempt to optimise near the existing settings.

## Set/Get parameters

Set or retrieve current alignment settings.

### Functions

- int digHoloConfigSetBeamCentre (int handleIdx, int axisIdx, int polIdx, float centre)
  *Sets the current beam centre in the camera plane for a specified axis (x/y) and polarisation component.*

- float digHoloConfigGetBeamCentre (int handleIdx, int axisIdx, int polIdx)
  *Returns the current beam centre in the camera plane for a specified axis (x/y) and polarisation component.*

- int digHoloConfigSetTilt (int handleIdx, int axisIdx, int polIdx, float tilt)
  *Sets the reference beam tilt in degrees. This corresponds with the position of the off-axis term in Fourier space that contains our field to be reconstructed.*

- float digHoloConfigGetTilt (int handleIdx, int axisIdx, int polIdx)
  *Gets the reference beam tilt in degrees. This corresponds with the position of the off-axis term in Fourier space that contains our field to be reconstructed.*

- int digHoloConfigSetDefocus (int handleIdx, int polIdx, float defocus)
  *Sets the the reference beam defocus in dioptre.*

- float digHoloConfigGetDefocus (int handleIdx, int polIdx)
  *Gets the reference beam defocus in dioptre.*

- int digHoloConfigSetPolLockTilt (int handleIdx, int polLock)
  *Sets whether the tilt parameter of the two polarisation components are independent (0) or locked to always have the value of the first polarisation (1).*

- int digHoloConfigGetPolLockTilt (int handleIdx)
  *Returns whether the tilt parameter of the two polarisation components are independent (0) or locked to always have the value of the first polarisation (1).*

- int digHoloConfigSetPolLockDefocus (int handleIdx, int polLock)
  *Sets whether the defocus parameter of the two polarisation components are independent (0) or locked to always have the value of the first polarisation (1).*

- int digHoloConfigGetPolLockDefocus (int handleIdx)
  *Returns whether the defocus parameter of the two polarisation components are independent (0) or locked to always have the value of the first polarisation (1).*

- int digHoloConfigSetPolLockBasisWaist (int handleIdx, int polLock)
  *Sets whether the waist parameter of the two polarisation components are independent (0) or locked to always have the value of the first polarisation (1).*

- int digHoloConfigGetPolLockBasisWaist (int handleIdx)

*Returns whether the waist parameter of the two polarisation components are independent (0) or locked to always have the value of the first polarisation (1).*

## Detailed Description

Set or retrieve current alignment settings.

## Function Documentation

### float digHoloConfigGetBeamCentre (int *handleIdx*, int *axisIdx*, int *polIdx*)

Returns the current beam centre in the camera plane for a specified axis (x/y) and polarisation component.

**Parameters**

| | | |
|---|---|---|
| in | *handleIdx* | : enumerated handle index of the digHoloObject |
| in | *axisIdx* | : x-axis (0), y-axis (1) |
| in | *polIdx* | : polarisation component (0,1) |

**Returns**

beamCentre : Current beam centre for specified axis and polarisation. Will return zero if any index is invalid (handle, axis or polarisation).

### float digHoloConfigGetDefocus (int *handleIdx*, int *polIdx*)

Gets the reference beam defocus in dioptre.

**Parameters**

| | | |
|---|---|---|
| in | *handleIdx* | : enumerated handle index of the digHoloObject |
| in | *polIdx* | : polarisation component [0,1] |

**Returns**

defocus : reference beam defocus in dioptre. Returns zero for invalid handle, axis or polarisation index.

### int digHoloConfigGetPolLockBasisWaist (int *handleIdx*)

Returns whether the waist parameter of the two polarisation components are independent (0) or locked to always have the value of the first polarisation (1).

**Parameters**

| | | |
|---|---|---|
| in | *handleIdx* | : enumerated handle index of the digHoloObject |

**Returns**

polLock : Independent polarisations (0), or locked to the value of the first polarisation (1).

**int digHoloConfigGetPolLockDefocus (int *handleIdx*)**

Returns whether the defocus parameter of the two polarisation components are independent (0) or locked to always have the value of the first polarisation (1).

When the PolLock is engaged, the defocus of the reference beam in both polarisation components is constrained to be the same. For example, if the optics of your digital holography setup are such that there's no physical reason why the curvature of the wavefront in H, and the tilt in V, are different, you could force them to be the same. Alternatively, if you know the defocus in H and V are similar, you can engage polarisation lock to reduce by 1 free parameter for initial optimisation, before letting the H and V have different tilts later for fine tweaking.

**Parameters**

| in | *handleIdx* | : enumerated handle index of the digHoloObject |
|----|-------------|------------------------------------------------|

**Returns**

polLock : Independent polarisations (0), or locked to the value of the first polarisation (1).

**int digHoloConfigGetPolLockTilt (int *handleIdx*)**

Returns whether the tilt parameter of the two polarisation components are independent (0) or locked to always have the value of the first polarisation (1).

When the PolLock is engaged, the tilt of the reference beam in both polarisation components is constrained to be the same. For example, if the optics of your digital holography setup are such that there's no physical reason why the tilt in H, and the tilt in V, are different, you could force them to be the same. Alternatively, if you know the tilt in H and V are similar, you can engage polarisation lock to reduce by 1 free parameter for initial optimisation, before letting the H and V have different tilts later for fine tweaking.

**Parameters**

| in | *handleIdx* | : enumerated handle index of the digHoloObject |
|----|-------------|------------------------------------------------|

**Returns**

polLock : Independent polarisations (0), or locked to the value of the first polarisation (1). Returns zero for invalid handle index.

**float digHoloConfigGetTilt (int *handleIdx*, int *axisIdx*, int *polIdx*)**

Gets the reference beam tilt in degrees. This corresponds with the position of the off-axis term in Fourier space that contains our field to be reconstructed.

**Parameters**

| in | *handleIdx* | : enumerated handle index of the digHoloObject |
|----|-------------|------------------------------------------------|
| in | *axisIdx*   | : x-axis (0), y-axis (1)                       |
| in | *polIdx*    | : polarisation component [0,1]                 |

**Returns**

tilt : angle of reference beam in degrees. Returns zero for invalid handle, axis or polarisation index.

**int digHoloConfigSetBeamCentre (int *handleIdx*, int *axisIdx*, int *polIdx*, float *centre*)**

Sets the current beam centre in the camera plane for a specified axis (x/y) and polarisation component.

**Parameters**

| | | |
|---|---|---|
| in | *handleIdx* | : enumerated handle index of the digHoloObject |
| in | *axisIdx* | : x-axis (0), y-axis (1) |
| in | *polIdx* | : polarisation component (0,1) |
| in | *centre* | : beam centre value |

**Returns**

errorCode : [DIGHOLO_ERROR_SUCCESS DIGHOLO_ERROR_INVALIDHANDLE DIGHOLO_ERROR_INVALIDAXIS DIGHOLO_ERROR_INVALIDPOLARISATION DIGHOLO_ERROR_INVALIDARGUMENT]

### int digHoloConfigSetDefocus (int *handleIdx*, int *polIdx*, float *defocus*)

Sets the the reference beam defocus in dioptre.

**Parameters**

| | | |
|---|---|---|
| in | *handleIdx* | : enumerated handle index of the digHoloObject |
| in | *polIdx* | : polarisation component [0,1] |
| in | *defocus* | : angle of reference beam in degrees. |

**Returns**

errorCode : [DIGHOLO_ERROR_SUCCESS DIGHOLO_ERROR_INVALIDHANDLE DIGHOLO_ERROR_INVALIDPOLARISATION DIGHOLO_ERROR_INVALIDARGUMENT]

### int digHoloConfigSetPolLockBasisWaist (int *handleIdx*, int *polLock*)

Sets whether the waist parameter of the two polarisation components are independent (0) or locked to always have the value of the first polarisation (1).

**Parameters**

| | | |
|---|---|---|
| in | *handleIdx* | : enumerated handle index of the digHoloObject |
| in | *polLock* | : Independent polarisations (0), or locked to the value of the first polarisation (1). |

**Returns**

errorCode : [DIGHOLO_ERROR_SUCCESS DIGHOLO_ERROR_INVALIDHANDLE]

### int digHoloConfigSetPolLockDefocus (int *handleIdx*, int *polLock*)

Sets whether the defocus parameter of the two polarisation components are independent (0) or locked to always have the value of the first polarisation (1).

The locking of the polarisations is not enforced immediately, it is enforced during frame processing.

When the PolLock is engaged, the defocus of the reference beam in both polarisation components is constrained to be the same. For example, if the optics of your digital holography setup are such that there's no physical reason why the curvature of the wavefront in H, and the tilt in V, are different, you could force them to be the same. Alternatively, if you know the defocus in H and V are similar, you can engage polarisation lock to reduce by 1 free parameter for initial optimisation, before letting the H and V have different tilts later for fine tweaking.

**Parameters**

| in | *handleIdx* | : enumerated handle index of the digHoloObject |
|---|---|---|
| in | *polLock* | : Independent polarisations (0), or locked to the value of the first polarisation (1). |

**Returns**

errorCode : [DIGHOLO_ERROR_SUCCESS DIGHOLO_ERROR_INVALIDHANDLE]

### int digHoloConfigSetPolLockTilt (int *handleIdx*, int *polLock*)

Sets whether the tilt parameter of the two polarisation components are independent (0) or locked to always have the value of the first polarisation (1).

The locking of the polarisations is not enforced immediately, it is enforced during frame processing.

When the PolLock is engaged, the tilt of the reference beam in both polarisation components is constrained to be the same. For example, if the optics of your digital holography setup are such that there's no physical reason why the tilt in H, and the tilt in V, are different, you could force them to be the same. Alternatively, if you know the tilt in H and V are similar, you can engage polarisation lock to reduce by 1 free parameter for initial optimisation, before letting the H and V have different tilts later for fine tweaking.

**Parameters**

| in | *handleIdx* | : enumerated handle index of the digHoloObject |
|---|---|---|
| in | *polLock* | : Independent polarisations (0), or locked to the value of the first polarisation (1). |

**Returns**

errorCode : [DIGHOLO_ERROR_SUCCESS DIGHOLO_ERROR_INVALIDHANDLE]

### int digHoloConfigSetTilt (int *handleIdx*, int *axisIdx*, int *polIdx*, float *tilt*)

Sets the reference beam tilt in degrees. This corresponds with the position of the off-axis term in Fourier space that contains our field to be reconstructed.

**Parameters**

| in | *handleIdx* | : enumerated handle index of the digHoloObject |
|---|---|---|
| in | *axisIdx* | : x-axis (0), y-axis (1) |
| in | *polIdx* | : polarisation component [0,1] |
| in | *tilt* | : angle of reference beam in degrees. |

**Returns**

errorCode : [DIGHOLO_ERROR_SUCCESS DIGHOLO_ERROR_INVALIDHANDLE DIGHOLO_ERROR_INVALIDPOLARISATION DIGHOLO_ERROR_INVALIDARGUMENT]

# Modal decomposition

Routines for configuring the spatial modal basis (if any) (e.g. Hermite-Gaussian, Laguerre-Gaussian)

## Modules

- Basis types
  *Supported types of modal spatial basis. e.g. Hermite-Gaussian, Laguerre-Gaussian, Custom.*
- Transfer matrix analysis parameters
  *Different metrics for assessing the quality of transfer matrix.*

## Functions

- int digHoloConfigSetBasisGroupCount (int handleIdx, int groupCount)
  *Sets the number of Hermite-Gaussian(m,n) modes to include in the basis. Less than groupCount=m+n.*

- int digHoloConfigGetBasisGroupCount (int handleIdx)
- int digHoloConfigSetBasisWaist (int handleIdx, int polIdx, float waist)
  *Sets the beam waist of the Hermite-Gaussian modes of which the reconstructed fields are composed (when maxMG>0)*

- float digHoloConfigGetBasisWaist (int handleIdx, int polIdx)
  *Returns the beam waist of the Hermite-Gaussian modes of which the reconstructed fields are composed (when maxMG>0)*

- int digHoloConfigSetBasisTypeHG (int handleIdx)
  *Sets the current modal basis to Hermite-Gaussian.*

- int digHoloConfigSetBasisTypeLG (int handleIdx)
  *Sets the current modal basis to Laguerre-Gaussian.*

- int digHoloConfigSetBasisTypeCustom (int handleIdx, int modeCountIn, int modeCountOut, complex64 *transform)
  *Sets the current modal basis to a custom basis.*

- int digHoloConfigSetBasisType (int handleIdx, int basisType)
- int digHoloConfigGetBasisType (int handleIdx)

## Detailed Description

Routines for configuring the spatial modal basis (if any) (e.g. Hermite-Gaussian, Laguerre-Gaussian)

## Function Documentation

### int digHoloConfigGetBasisGroupCount (int *handleIdx*)

Gets the number of Hermite-Gaussian(m,n) modes to include in the basis. Less than groupCount=m+n e.g. groupCount=1 would including only the fundamental mode. groupCount=0 would include HG no modal decomposition.

**Parameters**

| in | *handleIdx* | : enumerated handle index of the digHoloObject |
|---|---|---|

**Returns**

groupCount : Hermite-Gaussian modes less than order groupCount are included in the basis. Returns zero for invalid handle index.

### int digHoloConfigGetBasisType (int *handleIdx*)

Returns the current modal basis type [0: HG, 1: LG, 2: Custom]

**Parameters**

| in | *handleIdx* | : enumerated handle index of the digHoloObject |
|---|---|---|

**Returns**

basisType : 0: Hermite-Gaussian, 1: Laguerre-Gaussian, 2: Custom. Returns zero for invalid handle index.

### float digHoloConfigGetBasisWaist (int *handleIdx*, int *polIdx*)

Returns the beam waist of the Hermite-Gaussian modes of which the reconstructed fields are composed (when maxMG>0)

**Warning**

Separate waists for each polarisation component is not currently implemented.

**Parameters**

| in | *handleIdx* | : enumerated handle index of the digHoloObject |
|---|---|---|
| in | *polIdx* | : index of polarisation component |

**Returns**

waist : beam waist. Will return zero if an invalid handle index or polarisation index is supplied.

### int digHoloConfigSetBasisGroupCount (int *handleIdx*, int *groupCount*)

Sets the number of Hermite-Gaussian(m,n) modes to include in the basis. Less than groupCount=m+n.

e.g. groupCount=1 would including only the fundamental mode. groupCount=0 would include HG no modal decomposition. Formerly called digHoloConfigSetMaxMG

**Parameters**

| in | *handleIdx* | : enumerated handle index of the digHoloObject |
|---|---|---|
|  | *groupCount* | : Hermite-Gaussian modes less than order groupCount will be included in the basis. |

**Returns**

errorCode : [DIGHOLO_ERROR_SUCCESS DIGHOLO_ERROR_INVALIDHANDLE]

### int digHoloConfigSetBasisType (int *handleIdx*, int *basisType*)

Sets the current modal basis to a the specified type [0: HG, 1: LG, 2: Custom]

For custom basis, a call to digHoloConfigSetBasisTypeCustom() must have been called prior to the next modal decomposition. Otherwise, basis will default back to HG.

**Parameters**

| in | *handleIdx* | : enumerated handle index of the digHoloObject |
|---|---|---|
| in | *basisType* | : 0: Hermite-Gaussian, 1: Laguerre-Gaussian, 2: Custom |

**Returns**

errorCode : [DIGHOLO_ERROR_SUCCESS, DIGHOLO_ERROR_INVALIDHANDLE, DIGHOLO_ERROR_INVALIDDIMENSION, DIGHOLO_ERROR_NULLPOINTER, DIGHOLO_ERROR_MEMORYALLOCATION]

### int digHoloConfigSetBasisTypeCustom (int *handleIdx*, int *modeCountIn*, int *modeCountOut*, complex64 * *transform*)

Sets the current modal basis to a custom basis.

Internally, all fields are natively decomposed in the Hermite-Gaussian (HG) basis, as the HG basis is separable in X and Y, making for more efficient computation. The HG modal coefficients are then transformed into the custom basis using a matrix multiplication by the supplied transform matrix.

The specified matrix can include more Hermite-Gaussian modes than BasisGroupCount supports, however that part of the matrix will not be used in the transform. The supplied transform matrix is copied into an internal buffer at the time this function is called. Hence, it is safe to deallocate 'transform' afterwards. This also means that if 'transform' is changed, either the pointer or the data itself, then this function will need to be called again in order for the matrix to be updated.

**Parameters**

| in | *handleIdx* | : enumerated handle index of the digHoloObject |
|---|---|---|
| in | *modeCountIn* | : number of input Hermite-Gaussian modes |
| in | *modeCountOut* | : number of output custom modes |
| in | *transform* | : pointer to a matrix of dimensions modeCountOut x modeCountIn that specifies the transformation. |

**Returns**

errorCode : [DIGHOLO_ERROR_SUCCESS, DIGHOLO_ERROR_INVALIDHANDLE, DIGHOLO_ERROR_INVALIDDIMENSION, DIGHOLO_ERROR_NULLPOINTER, DIGHOLO_ERROR_MEMORYALLOCATION]

### int digHoloConfigSetBasisTypeHG (int *handleIdx*)

Sets the current modal basis to Hermite-Gaussian.

**Parameters**

| in | *handleIdx* | : enumerated handle index of the digHoloObject |
|---|---|---|

**Returns**

errorCode : [DIGHOLO_ERROR_SUCCESS, DIGHOLO_ERROR_INVALIDHANDLE]

### int digHoloConfigSetBasisTypeLG (int *handleIdx*)

Sets the current modal basis to Laguerre-Gaussian.

Internally, all fields are natively decomposed in the Hermite-Gaussian (HG) basis, as the HG basis is separable in X and Y, making for more efficient computation. The HG modal coefficients are then transformed into the Laguerre-Gaussian (LG) basis using a matrix multiplication.

**Parameters**

| | | |
|---|---|---|
| in | *handleIdx* | : enumerated handle index of the digHoloObject |

**Returns**

errorCode : [DIGHOLO_ERROR_SUCCESS, DIGHOLO_ERROR_INVALIDHANDLE]

**int digHoloConfigSetBasisWaist (int *handleIdx*, int *polIdx*, float *waist*)**

Sets the beam waist of the Hermite-Gaussian modes of which the reconstructed fields are composed (when maxMG>0)

**Warning**

Different waist values for each polarisation component are currently not supported.

**Parameters**

| | | |
|---|---|---|
| in | *handleIdx* | : enumerated handle index of the digHoloObject |
| in | *polIdx* | : index of polarisation component |
| in | *waist* | : beam waist |

**Returns**

errorCode : [DIGHOLO_ERROR_SUCCESS DIGHOLO_ERROR_INVALIDHANDLE DIGHOLO_ERROR_INVALIDPOLARISATION DIGHOLO_ERROR_INVALIDDIMENSION]

# Basis types

Supported types of modal spatial basis. e.g. Hermite-Gaussian, Laguerre-Gaussian, Custom.

## Macros

- #define DIGHOLO_BASISTYPE_HG   0
- #define DIGHOLO_BASISTYPE_LG   1
- #define DIGHOLO_BASISTYPE_CUSTOM   2

## Detailed Description

Supported types of modal spatial basis. e.g. Hermite-Gaussian, Laguerre-Gaussian, Custom.

Defines the basis the overlap coefficients are defined in. Natively, everything is decomposed as HG modes, which is then transformed into another basis through a matrix multiplication. That is, the only basis that the fields can be directly decomposed in is the HG basis, with all other bases being derived from those HG coefficients by multiplication through an additional matrix transformation.

## Macro Definition Documentation

### #define DIGHOLO_BASISTYPE_CUSTOM   2

Custom. Defined by a user-supplied transform matrix which maps Hermite-Gaussian modes onto the desired custom basis.

### #define DIGHOLO_BASISTYPE_HG   0

Hermite-Gaussian

### #define DIGHOLO_BASISTYPE_LG   1

Laguerre-Gaussian. Currently only supported up to 100 mode groups. (m+n=2p+|l|<=100)

# Transfer matrix analysis parameters

Different metrics for assessing the quality of transfer matrix.

## Macros

- #define DIGHOLO_METRIC_IL   0
- #define DIGHOLO_METRIC_MDL   1
- #define DIGHOLO_METRIC_DIAG   2
- #define DIGHOLO_METRIC_SNRAVG   3
- #define DIGHOLO_METRIC_DIAGBEST   4
- #define DIGHOLO_METRIC_DIAGWORST   5
- #define DIGHOLO_METRIC_SNRBEST   6
- #define DIGHOLO_METRIC_SNRWORST   7
- #define DIGHOLO_METRIC_SNRMG   8
- #define DIGHOLO_METRIC_COUNT   9

## Detailed Description

Different metrics for assessing the quality of transfer matrix.

The digHoloConfigGetAutoAlignMetrics routine, when run after a digHoloAutoAlign, will return an array of properties of the transfer matrix after auto-alignment. These defines enumerate the elements of that array. At least one of these metrics would be used to guide the AutoAlign routine, but the user can also see other metrics as well if available.

## Macro Definition Documentation

### #define DIGHOLO_METRIC_COUNT   9

Number of different metrics the powermeter routine outputs (e.g. il,mdl, SNR etc.)

### #define DIGHOLO_METRIC_DIAG   2

Total power along the diagonal of the transfer matrix

### #define DIGHOLO_METRIC_DIAGBEST   4

Diagonal element with the most power

### #define DIGHOLO_METRIC_DIAGWORST   5

Diagonal element with the worst power

### #define DIGHOLO_METRIC_IL   0

Insertion loss

### #define DIGHOLO_METRIC_MDL   1

Mode dependent loss (a.k.a matrix condition number)

### #define DIGHOLO_METRIC_SNRAVG   3

Total power along the diagonal/total power off-diagonal

**#define DIGHOLO_METRIC_SNRBEST   6**

    Diagonal/off-diagonal power of the best batch element

**#define DIGHOLO_METRIC_SNRMG   8**

    Same as SNRAVG, except coupling within degenerate HG mode groups counts as signal

**#define DIGHOLO_METRIC_SNRWORST   7**

    Diagonal/off-diagonal power of the worst batch element

# Execute batch processing

Routines for running a processing batch using current settings.

## Functions

- complex64 * digHoloProcessBatch (int handleIdx, int *batchCount, int *modeCount, int *polCount)
  *Processes a batch of frames using the current settings, and returns a modal decomposition of the resulting fields.*

- complex64 * digHoloProcessBatchFrequencySweepLinear (int handleIdx, int *batchCount, int *modeCount, int *polCount, float lambdaStart, float lambdaStop, int lambdaCount)
  *Processes a batch of frames captured at different wavelengths using the current settings, and returns a modal decomposition of the resulting fields.*

- complex64 * digHoloProcessBatchWavelengthSweepArbitrary (int handleIdx, int *batchCount, int *modeCount, int *polCount, float *wavelengths, int lambdaCount)
  *Processes a batch of frames captured at different wavelengths using the current settings, and returns a modal decomposition of the resulting fields.*

## Detailed Description

Routines for running a processing batch using current settings.

## Function Documentation

### complex64 * digHoloProcessBatch (int *handleIdx*, int * *batchCount*, int * *modeCount*, int * *polCount*)

Processes a batch of frames using the current settings, and returns a modal decomposition of the resulting fields.

**Parameters**

| | | |
|---|---|---|
| in | *handleIdx* | : enumerated handle index of the digHoloObject |
| out | *batchCount* | :int32 pointer where the total number of output fields will be returned. Returns zero for invalid handle index. |
| out | *modeCount* | : int32 pointer where the total number of spatial modes will be returned. Returns zero for invalid handle index. |
| out | *polCount* | : int32 pointer where the total number of polarisation components per field will be returned. Returns zero for invalid handle index. |

**Returns**

basisCoefs : pointer to a complex64 containing a batchCount x (polCount x modeCount) array containing the modal coefficients for the reconstructed fields. If no modal basis is set (e.g. BasisGroupCount=0), a nullpointer will be returned and modeCount will be zero. Returns a nullpointer for invalid handle index.

**complex64 * digHoloProcessBatchFrequencySweepLinear (int   *handleIdx*, int * *batchCount*, int *   *modeCount*, int *   *polCount*, float   *lambdaStart*, float   *lambdaStop*, int   *lambdaCount*)**

Processes a batch of frames captured at different wavelengths using the current settings, and returns a modal decomposition of the resulting fields.

Similar to [digHoloProcessBatch()](), except you can specify that the frames are from a linear frequency sweep from lambaStart to lambdaStop in lambdaCount equally spaced frequency steps. If batchCount>lambdaCount, modular wrapping will be applied. For example, processing multiple frequency sweeps in a single processing batch.

**Parameters**

| in  | handleIdx   | : enumerated handle index of the digHoloObject |
|-----|-------------|------------------------------------------------|
| out | batchCount  | :int32 pointer where the total number of output fields will be returned. Returns zero for invalid handle index. |
| out | modeCount   | : int32 pointer where the total number of spatial modes will be returned. Returns zero for invalid handle index. |
| out | polCount    | : int32 pointer where the total number of polarisation components per field will be returned. Returns zero for invalid handle index. |
| in  | lambdaStart | : Wavelength at start of sweep. |
| in  | lambdaStop  | : Wavelength at end of sweep. |
| in  | lambdaCount | : Total number of frequency steps between lambdaStart and lambdaStop. |

**Returns**

basisCoefs : pointer to a complex64 containing a batchCount x (polCount x modeCount) array containing the modal coefficients for the reconstructed fields. If no modal basis is set (e.g. BasisGroupCount=0), a nullpointer will be returned and modeCount will be zero. Returns a nullpointer for invalid handle index.

**complex64 * digHoloProcessBatchWavelengthSweepArbitrary (int   *handleIdx*, int * *batchCount*, int *   *modeCount*, int *   *polCount*, float *   *wavelengths*, int *lambdaCount*)**

Processes a batch of frames captured at different wavelengths using the current settings, and returns a modal decomposition of the resulting fields.

Similar to [digHoloProcessBatch()](), except you can specify that the frames were captured at wavelengths as per a supplied array. If batchCount>lambdaCount, modular wrapping will be applied. For example, processing multiple frequency sweeps in a single processing batch.

**Parameters**

| in  | handleIdx   | : enumerated handle index of the digHoloObject |
|-----|-------------|------------------------------------------------|
| out | batchCount  | :int32 pointer where the total number of output fields will be returned. Returns zero for invalid handle index. |
| out | modeCount   | : int32 pointer where the total number of spatial modes will be returned. Returns zero for invalid handle index. |
| out | polCount    | : int32 pointer where the total number of polarisation components per field will be returned. Returns zero for invalid handle index. |
| in  | wavelengths | : pointer to array of wavelengths |
| in  | lambdaCount | : length of the wavelengths array. |

**Returns**

basisCoefs : pointer to a complex64 containing a batchCount x (polCount x modeCount) array containing the modal coefficients for the reconstructed fields. If no modal basis is set (e.g. BasisGroupCount=0), a nullpointer will be returned and modeCount will be zero. Returns a nullpointer for invalid handle index.

# Return reconstructed fields and modal coefficients

Returns the final reconstructed fields after digHolo processing.

## Functions

- complex64 * digHoloGetFields (int handleIdx, int *batchCount, int *polCount, float **x, float **y, int *width, int *height)
  *Return reconstructed fields and corresponding x and y axis.*

- int digHoloGetFields16 (int handleIdx, int *batchCount, int *polCount, short **fieldR, short **fieldI, float **fieldScale, float **x, float **y, int *width, int *height)
- complex64 * digHoloBasisGetFields (int handleIdx, int *batchCount, int *polCount, float **x, float **y, int *width, int *height)
- complex64 * digHoloBasisGetCoefs (int handleIdx, int *batchCount, int *modeCount, int *polCount)
  *Returns a pointer to an array containing the modal coefficients for the latest processed batch.*

## Detailed Description

Returns the final reconstructed fields after digHolo processing.

## Function Documentation

### complex64 * digHoloBasisGetCoefs (int *handleIdx*, int * *batchCount*, int * *modeCount*, int * *polCount*)

Returns a pointer to an array containing the modal coefficients for the latest processed batch.

Array will be of size batchCount x (poltCount x modeCount). The mode coefficients for each polarisation component are stored as separate blocks of memory. e.g. HHHHHVVVVV, not HVHVHVHV The returned array is the same as for the digHoloProcessBatch*() routines, except this routine does not trigger any processing.

**Parameters**

| in | *handleIdx* | : enumerated handle index of the digHoloObject |
|---|---|---|
| out | *batchCount* | :int32 pointer where the total number of output fields will be returned. Returns zero for invalid handle index. |
| out | *modeCount* | : int32 pointer where the total number of spatial modes will be returned. Returns zero for invalid handle index. |
| out | *polCount* | : int32 pointer where the total number of polarisation components per field will be returned. Returns zero for invalid handle index. |

**Returns**

basisCoefs : pointer to a complex64 containing a batchCount x (polCount modeCount) array containing the modal coefficients for the reconstructed fields. If no modal basis is set (e.g. BasisGroupCount=0) or no processing has been previously performed, a nullpointer will be returned and modeCount will be zero. Returns a nullpointer for invalid handle index.

**complex64 * digHoloBasisGetFields (int   *handleIdx*, int \*   *batchCount*, int \* *polCount*, float \*\*   *x*, float \*\*   *y*, int \*   *width*, int \*   *height*)**

Return fields of the current basis and corresponding x and y axis.

Basis fields are not normally stored in 2D in float32 format, so they are constructed specifically for export when this function is called.

Returns pointer to internal memory storing the fields. The fields are actually stored internally as int16 in separable 1D format. Hence this routine construct a float32 complex version of the field and overwrites it to an internal complex buffer used as part of the processing pipeline.

**Parameters**

| | | |
|---|---|---|
| in | *handleIdx* | : enumerated handle index of the digHoloObject |
| out | *batchCount* | : pointer to an int32 where the number of fields in the basis per polarisation will be returned. |
| out | *polCount* | : pointer to an int32 where the number of polarisation components will be returned. |
| out | *x* | : pointer to an array containing the x-axis of the field. |
| out | *y* | : pointer to an array containing the y-axis of the field. |
| out | *width* | : pointer to an int32 where the length of the x-axis of the field per polarisation will be returned. |
| out | *height* | : pointer to an int32 where the length of the y-axis of the field per polarisation will be returned. |

**Returns**

fields : batchCount x polCount x width x height complex64 array containing the fields

**complex64 * digHoloGetFields (int   *handleIdx*, int \*   *batchCount*, int \*   *polCount*, float \*\*   *x*, float \*\*   *y*, int \*   *width*, int \*   *height*)**

Return reconstructed fields and corresponding x and y axis.

Returns pointer to internal memory storing the fields. The fields are actually stored internally as int16 (see digHoloGetFields16). Hence this routine construct a float32 complex version of the field and overwrites it to an internal complex buffer used as part of the processing pipeline.

**Parameters**

| | | |
|---|---|---|
| in | *handleIdx* | : enumerated handle index of the digHoloObject |
| out | *batchCount* | : pointer to an int32 where the number of fields in the batch per polarisation will be returned. |
| out | *polCount* | : pointer to an int32 where the number of polarisation components will be returned. |
| out | *x* | : pointer to an array containing the x-axis of the field. |
| out | *y* | : pointer to an array containing the y-axis of the field. |
| out | *width* | : pointer to an int32 where the length of the x-axis of the field per polarisation will be returned. |
| out | *height* | : pointer to an int32 where the length of the y-axis of the field per polarisation will be returned. |

**Returns**

fields : batchCount x polCount x width x height complex64 array containing the fields

**int digHoloGetFields16 (int   *handleIdx*, int \*   *batchCount*, int \*   *polCount*, short \*\*   *fieldR*, short \*\*   *fieldI*, float \*\*   *fieldScale*, float \*\*   *x*, float \*\*   *y*, int \*   *width*, int \*   *height*)**

Return reconstructed fields and corresponding x and y axis in int16 format.

Returns pointers to the internal memory storing the fields. In order to convert to a typical complex 32-bit float field, you'd have to divide (fieldR+1i*fieldI)/fieldScale for each field component Hence this routine construct a float32 complex version of the field and overwrites it to an internal complex buffer used as part of the processing pipeline.

**Parameters**

| | | |
|---|---|---|
| in | *handleIdx* | : enumerated handle index of the digHoloObject |
| out | *batchCount* | : pointer to an int32 where the number of fields in the batch per polarisation will be returned. |
| out | *polCount* | : pointer to an int32 where the number of polarisation components will be returned. |
| out | *fieldR* | : pointer to an array containing the int16 real components of the field. (batchCount x polCount) |
| out | *fieldI* | : pointer to an array containing the int16 imag components of the field. (batchCount x polCount) |
| out | *fieldScale* | : a pointer to an array containg the float32 division factors to apply, (fieldR+1i.*fieldI)/fieldScale, in order to convert the field to it's proper representation. |
| out | *x* | : pointer to an array containing the x-axis of the field. |
| out | *y* | : pointer to an array containing the y-axis of the field. |
| out | *width* | : pointer to an int32 where the length of the x-axis of the field per polarisation will be returned. |
| out | *height* | : pointer to an int32 where the length of the y-axis of the field per polarisation will be returned. |

**Returns**

errorCode : [DIGHOLO_ERROR_SUCCESS, DIGHOLO_ERROR_INVALIDHANDLE, DIGHOLO_ERROR_NULLPOINTER]

# Advanced functions

Provides access to internal buffers, batch summaries and individual internal steps in the processing pipeline.

## Modules
- [Analysis parameter types](#)
  *Indices of the parameters that are stored in the analysis array.*

## Functions
- int [digHoloBatchGetSummary](#) (int handleIdx, int planeIdx, int *parameterCount, int *totalCount, int *polCount, float **parameters, int *pixelCountX, int *pixelCountY, float **xAxis, float **yAxis, float **totalIntensity)
  *Returns parameters that analyse and summarise the frames and fields from a batch process.*

- int [digHoloProcessFFT](#) (int handleIdx)
  *Runs the FFT step of the processing pipeline.*

- int [digHoloProcessIFFT](#) (int handleIdx)
  *Runs the IFFT step of the processing pipeline.*

- int [digHoloProcessRemoveTilt](#) (int handleIdx)
  *Runs the reference beam tilt removal step of the processing pipeline.*

- int [digHoloProcessBasisExtractCoefs](#) (int handleIdx)
  *This step overlaps the reconstructed field with each of the basis modes and extracts the corresponding complex coefficients.*

- complex64 * [digHoloGetFourierPlaneFull](#) (int handleIdx, int *batchCount, int *polCount, int *width, int *height)
  *Returns the buffer and the associated dimensions of the real-to-complex Fourier transform of the FrameBuffer. This buffer will not exist until an FFT is performed.*

- complex64 * [digHoloGetFourierPlaneWindow](#) (int handleIdx, int *batchCount, int *polCount, int *width, int *height)
  *Returns the selected portion of Fourier space that was IFFT'd to become the reconstructed field.*

## Detailed Description
Provides access to internal buffers, batch summaries and individual internal steps in the processing pipeline.

## Function Documentation

**int digHoloBatchGetSummary (int   *handleIdx*, int   *planeIdx*, int *   *parameterCount*, int *   *totalCount*, int *   *polCount*, float **   *parameters*, int *   *pixelCountX*, int *   *pixelCountY*, float **   *xAxis*, float **   *yAxis*, float **   *totalIntensity*)**

Returns parameters that analyse and summarise the frames and fields from a batch process.

At the end of the internal FFT routine, as well as during the ApplyTilt routine which is typically run after the IFFT routine, all fields are analysed to extract properties like total power, effective area, centre of mass etc. The types of parameters can be seen as the defines DIGHOLO_ANALYSIS_* For each plane (Fourier plane of the camera, or reconstructed field plane), and each batch element, and each polarisation, these properties are calculated, as well as averages over the entire batch.

The returned 'parameters' array is of dimensions parameterCount x polCount x totalCount

Warning : MAXIDX parameter is actually an int32, not a float32. Hence when reading this parameter, you must interpret the 32 bits as int32 not float32 format.

**Parameters**

| in  | *handleIdx*      | : enumerated handle index of the digHoloObject |
|-----|------------------|-----------------------------------------------|
| in  | *planeIdx*       | : Fourier plane of camera (0), Reconstructed field plane (1). i.e. the result of the FFT (0), and the result of the IFFT after tilt removal (1) |
| out | *parameterCount* | : number of parameters (DIGHOLO_ANALYSIS_COUNT) |
| out | *totalCount*     | : the total number of calculated values for each parameter in each polarisation. This would be the batchCount x avgCount +1. In the Fourier plane (planeIdx=0), every Fourier transformed frame (batchCount x avgCount) has parameters calculated for it, plus an additional calculation for the average over all. For the reconstructed field plane (planeIdx=1), every Inverse Fourier transformed element has parameters calculated for it (batchCount), plus 1 for the average over all batch elements. If averaging is employed, the array dimensions for the planeIdx=1 remains the same as for planeIdx=0 (batchCount x avgCount +1), however only the first batchCount+1 elements will be meaningful. |
| out | *polCount*       | : the number of polarisation components per totalCount per parameterCount |
| out | *parameters*     | : pointer to array containing the parameters (parameterCount x polCount x totalCount) |
| out | *pixelCountX*    | : pointer to an int32 into which the x dimension of the totalIntensity will be returned |
| out | *pixelCountY*    | : pointer to an int32 into which the y dimension of the totalIntensity will be returned |
| out | *xAxis*          | : a pointer to a pointer into which the address of an array containing the x-axis of the plane will be returned |
| out | *yAxis*          | : a pointer to a pointer into which the address of an array containing the y-axis of the plane will be returned |
| out | *totalIntensity* | : the sum of the intensity over all batch elements in the specified plane. |

**Returns**

errorCode : [DIGHOLO_ERROR_SUCCESS, DIGHOLO_ERROR_INVALIDHANDLE, DIGHOLO_ERROR_INVALIDDIMENSION, DIGHOLO_ERROR_NULLPOINTER]

**complex64 * digHoloGetFourierPlaneFull (int   *handleIdx*, int *   *batchCount*, int *   *polCount*, int *   *width*, int *   *height*)**

Returns the buffer and the associated dimensions of the real-to-complex Fourier transform of the FrameBuffer. This buffer will not exist until an FFT is performed.

**Parameters**

| | | |
|---|---|---|
| in | *handleIdx* | : enumerated handle index of the digHoloObject |
| out | *batchCount* | : a pointer to an int32 where the number of Fourier planes per polarisation will be returned. |
| out | *polCount* | : a pointer to an int32 where the number of polarisation components will be returned. |
| out | *width* | : a pointer to an int32 where the x-axis dimension of the Fourier plane will be returned. Remember, this is a real-to-complex transform, so the size will be {(w/2+1) x h}, rather than {w x h} of a full transform. |
| out | *height* | : a pointer to an int32 where the y-axis dimension of the Fourier plane will be returned. |

**Returns**

 fourierPlanes : a pointer to a batchCount x polCount x width x height array containing the Fourier planes from the last batch of Fourier processing. Will return null pointer for invalid handle index.

**complex64 * digHoloGetFourierPlaneWindow (int *handleIdx*, int * *batchCount*, int * *polCount*, int * *width*, int * *height*)**

Returns the selected portion of Fourier space that was IFFT'd to become the reconstructed field.

This buffer will not exist until after an IFFT is performed, as the window is selected only at the start of the IFFT processing step.

**Parameters**

| | | |
|---|---|---|
| in | *handleIdx* | : enumerated handle index of the digHoloObject |
| out | *batchCount* | : a pointer to an int32 where the number of Fourier windows per polarisation will be returned. |
| out | *polCount* | : a pointer to an int32 where the number of polarisation components will be returned. |
| out | *width* | : a pointer to an int32 where the x-axis dimension of the Fourier window will be returned. |
| out | *height* | : a pointer to an int32 where the y-axis dimension of the Fourier window will be returned. |

**Returns**

 fourierPlanes : a pointer to a batchCount x polCount x width x height array containing the Fourier windows from the last batch of IFT processing. Will return null pointer for invalid handle index.

**int digHoloProcessBasisExtractCoefs (int *handleIdx*)**

This step overlaps the reconstructed field with each of the basis modes and extracts the corresponding complex coefficients.

The resulting coefficients would be accessible via the digHoloBasisGetCoefs() routine. The basis fields would be accessible via the digHoloBasisGetFields() routine.

**Parameters**

| | | |
|---|---|---|
| in | *handleIdx* | : enumerated handle index of the digHoloObject |

**Returns**

    errorCode : [DIGHOLO_ERROR_SUCCESS, DIGHOLO_ERROR_INVALIDHANDLE, DIGHOLO_ERROR_NULLPOINTER]. If there is no basis (BasisGroupCount=0), this function will always return SUCCESS.

### int digHoloProcessFFT (int *handleIdx*)

Runs the FFT step of the processing pipeline.

This is the first step in the processing pipeline, which starts from the frameBuffer and performs the required Fourier transforms, as well as some additional analysis of the resulting Fourier planes for each transform.

**Parameters**

| in | *handleIdx* | : enumerated handle index of the digHoloObject |
|----|-------------|------------------------------------------------|

**Returns**

    errorCode : [DIGHOLO_ERROR_SUCCESS, DIGHOLO_ERROR_INVALIDHANDLE, DIGHOLO_ERROR_NULLPOINTER]

### int digHoloProcessIFFT (int *handleIdx*)

Runs the IFFT step of the processing pipeline.

This step selects the appropriate window in the Fourier plane and inverse Fourier transforms back into the plane of the camera creating the reconstructed field, which at this stage will still have some residual tilt due to the reference beam.

The Fourier window that was inverse Fourier transformed would be accessible using the digHoloGetFourierPlaneWindow() routine.

**Parameters**

| in | *handleIdx* | : enumerated handle index of the digHoloObject |
|----|-------------|------------------------------------------------|

**Returns**

    errorCode : [DIGHOLO_ERROR_SUCCESS, DIGHOLO_ERROR_INVALIDHANDLE, DIGHOLO_ERROR_NULLPOINTER]

### int digHoloProcessRemoveTilt (int *handleIdx*)

Runs the reference beam tilt removal step of the processing pipeline.

This step takes the reconstructed field output from the IFT and removes the residual tilt of the reference beam, as well as converting the field from float32 to int16 format. After this step, the field is fully reconstructed.

The resulting reconstructed field would be accessible using the digHoloGetFields() routine.

**Parameters**

| in | *handleIdx* | : enumerated handle index of the digHoloObject |
|----|-------------|------------------------------------------------|

**Returns**

    errorCode : [DIGHOLO_ERROR_SUCCESS, DIGHOLO_ERROR_INVALIDHANDLE, DIGHOLO_ERROR_NULLPOINTER]

# Analysis parameter types

Indices of the parameters that are stored in the analysis array.

## Macros

- #define DIGHOLO_ANALYSIS_TOTALPOWER   0
- #define DIGHOLO_ANALYSIS_COMX   1
- #define DIGHOLO_ANALYSIS_COMY   2
- #define DIGHOLO_ANALYSIS_MAXABS   3
- #define DIGHOLO_ANALYSIS_MAXABSIDX   4
- #define DIGHOLO_ANALYSIS_AEFF   5
- #define DIGHOLO_ANALYSIS_COMYWRAP   6
- #define DIGHOLO_ANALYSIS_COUNT   7

## Detailed Description

Indices of the parameters that are stored in the analysis array.

When a batch is processed, information about the processed frames are calculated in both the Fourier plane, and the reconstructed field plane. The information is stored in a 4D array (parameterIdx x planeIdx (0 or 1 for Fourier or reconstructed) x polarisationIdx x (total batch count, including averaging + 1 for the sum over the whole batch)

## Macro Definition Documentation

### #define DIGHOLO_ANALYSIS_AEFF   5

Effective area (Petermann II)

### #define DIGHOLO_ANALYSIS_COMX   1

centre of mass (x)

### #define DIGHOLO_ANALYSIS_COMY   2

centre of mass (y)

### #define DIGHOLO_ANALYSIS_COMYWRAP   6

centre of mass (y) if we assume the y-axis is wrapping around. Can be useful if you're working at the edge of Fourier space, and the off-axis term is wrapping around to the other side of Fourier space. i.e. you're working up near the absolute limit of reference beam angle. Possibly because the field you're trying to reconstruct is near the limit of your camera resolution.

### #define DIGHOLO_ANALYSIS_COUNT   7

Total count of analysis types

### #define DIGHOLO_ANALYSIS_MAXABS   3

maximum absolute value

### #define DIGHOLO_ANALYSIS_MAXABSIDX   4

index corresponding to ANALYSIS_MAXABS. Warning this is an int32, not a float32

**#define DIGHOLO_ANALYSIS_TOTALPOWER 0**
    Total power

# Multi-threading, performance and benchmarking

Routines for configuring multi-threading, performance and benchmarking performance.

## Modules

- Benchmark info
  *Enumeration of performance (Hz) for each subroutine returned by the digHoloBenchmark 'info' array.*
- FFTW library configuration
  *FFTW library configuration such as planning mode, wisdom file location etc.*

## Functions

- int digHoloConfigSetThreadCount (int handleIdx, int threadCount)
  *Sets the number of threads to be used for processing.*

- int digHoloConfigGetThreadCount (int handleIdx)
  *Gets the number of threads to be used for processing.*

- int digHoloBenchmarkEstimateThreadCountOptimal (int handleIdx, float goalDuration)
  *Runs full batch processing using current settings using threads from 1 to {maximum thread count, typically 2xlogical CPU count} Can be used to estimate the optimal number of threads for your machine. The answer would typically be ~the logical CPU count, or physical CPU count.*

- float digHoloBenchmark (int handleIdx, float goalDuration, float *info)
  *A benchmarking routine which will perform batch processing using the current settings and output performance profiling information about the various parts of the processing pipeline.*

## Detailed Description

Routines for configuring multi-threading, performance and benchmarking performance.

## Function Documentation

### float digHoloBenchmark (int  *handleIdx*, float  *goalDuration*, float *  *info*)

A benchmarking routine which will perform batch processing using the current settings and output performance profiling information about the various parts of the processing pipeline.

**Parameters**

| | | |
|---|---|---|
| in | *handleIdx* | : enumerated handle index of the digHoloObject |
| in | *goalDuration* | : The approximate total runtime for this benchmark routine (seconds). Longer duration = more accurate. |
| in | *info* | : An optional array of floats into which the performance (Hz) of each of the subroutines is written. |

**Returns**
> batchProcessingRate : the number of batches per second that can be processed using the current settings.

## int digHoloBenchmarkEstimateThreadCountOptimal (int *handleIdx*, float *goalDuration*)

Runs full batch processing using current settings using threads from 1 to {maximum thread count, typically 2xlogical CPU count} Can be used to estimate the optimal number of threads for your machine. The answer would typically be ~the logical CPU count, or physical CPU count.

**Parameters**

| in | *handleIdx* | : enumerated handle index of the digHoloObject |
|----|-------------|------------------------------------------------|
| in | *goalDuration* | : The approximate total runtime for this benchmark routine (seconds). Longer duration = more accurate. |

**Returns**
> threadCount : The recommended optimal thread count to use for the current processing settings.

## int digHoloConfigGetThreadCount (int *handleIdx*)

Gets the number of threads to be used for processing.

**Parameters**

| in | *handleIdx* | : enumerated handle index of the digHoloObject |
|----|-------------|------------------------------------------------|

**Returns**
> threadCount : number of threads to be used for processing

## int digHoloConfigSetThreadCount (int *handleIdx*, int *threadCount*)

Sets the number of threads to be used for processing.

**Parameters**

| in | *handleIdx* | : enumerated handle index of the digHoloObject |
|----|-------------|------------------------------------------------|
| in | *threadCount* | : number of threads to be used for processing |

**Returns**
> errorCode : [DIGHOLO_ERROR_SUCCESS, DIGHOLO_ERROR_INVALIDHANDLE]. If an invalid number of threads is requested. Less than one, or more than an internal THREADCOUNT_MAX, the threadCount will be set to defaults, but will not raise an error.

# Benchmark info

Enumeration of performance (Hz) for each subroutine returned by the digHoloBenchmark 'info' array.

## Macros

- #define DIGHOLO_BENCHMARK_FFT   0
- #define DIGHOLO_BENCHMARK_IFFT   1
- #define DIGHOLO_BENCHMARK_APPLYTILT   2
- #define DIGHOLO_BENCHMARK_BASIS   3
- #define DIGHOLO_BENCHMARK_OVERLAP   4
- #define DIGHOLO_BENCHMARK_TOTAL   5

## Detailed Description

Enumeration of performance (Hz) for each subroutine returned by the digHoloBenchmark 'info' array.

## Macro Definition Documentation

### #define DIGHOLO_BENCHMARK_APPLYTILT   2

The number of digHoloApplyTilt() routine calls per second

### #define DIGHOLO_BENCHMARK_BASIS   3

The rate at which the entire HG modal basis can be calculated (Hz)

### #define DIGHOLO_BENCHMARK_FFT   0

The number of digHoloFFT() routine calls per second.

### #define DIGHOLO_BENCHMARK_IFFT   1

The number of digHoloIFFT() routine calls per second

### #define DIGHOLO_BENCHMARK_OVERLAP   4

The number of digHoloOverlapModes() routine calls per second.

### #define DIGHOLO_BENCHMARK_TOTAL   5

The total number of frames per second that can be processed through the entire processing pipeline from digHoloFFT() to digHoloOverlapModes()

# FFTW library configuration

FFTW library configuration such as planning mode, wisdom file location etc.

## Functions

- int digHoloConfigSetFFTWPlanMode (int handleIdx, int planningMode)
  *Sets the FFTW planning mode used.*

- int digHoloConfigGetFFTWPlanMode (int handleIdx)
  *Gets the FFTW planning mode used.*

- int digHoloFFTWWisdomForget ()
  *Causes all accumulate FFTW wisdom to be discarded. https://www.fftw.org/fftw3_doc/Forgetting-Wisdom.html.*

- int digHoloFFTWWisdomFilename (const char *filename)
  *Changes the file used to store FFTW wisdom from the default, to the specified filename.*

## Detailed Description

FFTW library configuration such as planning mode, wisdom file location etc.

## Function Documentation

### int digHoloConfigGetFFTWPlanMode (int *handleIdx*)

Gets the FFTW planning mode used.

Internally, the FFTW library (www.fftw.org) is used. This function allows the FFTW planning mode to be selected. https://www.fftw.org/fftw3_doc/Planner-Flags.html If your program appears to freeze near the start of processing the first time it is run, it may be because the plan mode is set to 'exhaustive'.

**Parameters**

| in | *handleIdx* | : enumerated handle index of the digHoloObject |
|----|-------------|------------------------------------------------|

**Returns**

planningMode : 0: Estimate, 1: Measure, 2: Patient, 3: Exhaustive. Returns zero if an invalid handle index is supplied.

### int digHoloConfigSetFFTWPlanMode (int *handleIdx*, int *planningMode*)

Sets the FFTW planning mode used.

Internally, the FFTW library (www.fftw.org) is used. This function allows the FFTW planning mode to be selected. https://www.fftw.org/fftw3_doc/Planner-Flags.html If your program appears to freeze near the start of processing the first time it is run, it may be because the plan mode is set to 'exhaustive'.

**Parameters**

| | | |
|---|---|---|
| in | *handleIdx* | : enumerated handle index of the digHoloObject |
| in | *planningMode* | : 0: Estimate, 1: Measure, 2: Patient, 3: Exhaustive |

**Returns**

errorCode : [DIGHOLO_ERROR_SUCCESS, DIGHOLO_ERROR_INVALIDHANDLE DIGHOLO_ERROR_INVALIDARGUMENT]

### int digHoloFFTWWisdomFilename (const char * *filename*)

Changes the file used to store FFTW wisdom from the default, to the specified filename.

Nothing happens immediately, the filename is simply updated for use the next time FFTW wisdom is loaded/saved. https://www.fftw.org/doc/Words-of-Wisdom_002dSaving-Plans.html

**Returns**

errorCode : [DIGHOLO_ERROR_SUCCESS]

### int digHoloFFTWWisdomForget ()

Causes all accumulate FFTW wisdom to be discarded. https://www.fftw.org/fftw3_doc/Forgetting-Wisdom.html.

**Returns**

errorCode : [DIGHOLO_ERROR_SUCCESS]

## Diagnostics, console printing, debugging

Configuring how console printing is performed, plotting tools, debug tools, as well as additional utilities.

### Modules

- Viewport display modes
  *Types of plotting functions the GetViewport function supports. Mostly used for visually internal buffers such as the FourierPlane, reconstructed fields, etc.*
- Console printout verbosity levels
  *Specifies the desired level of detail which should be printed to the console.*

### Functions

- unsigned char * digHoloGetViewport (int handleIdx, int displayMode, int forceProcessing, int *width, int *height, char **windowString)
  *Visualises internal buffers. Renders an RGB bitmap based on what type of plot you want to see (displayMode).*

- int digHoloGetViewportToFile (int handleIdx, int displayMode, int forceProcessing, int *width, int *height, char **windowString, const char *filename)
  *Outputs the RGB pixel buffer returned by digHoloGetViewport to a binary file.*

- int digHoloConfigSetVerbosity (int handleIdx, int verbosity)
  *Specifies how much information is printed to console/file during operation.*

- int digHoloConfigGetVerbosity (int handleIdx)
  *Returns how much information is printed to console/file during operation.*

- int digHoloConsoleRedirectToFile (char *filename)
  *Calling this function will redirect stdout console to the specified filename.*

- int digHoloConsoleRestore ()
  *Sets the console to stdout.*

- int digHoloRunBatchFromConfigFile (char *filename)
  *Loads settings from a tab-delimited text file and performs a batch processing run.*

- int digHoloConfigBackupSave (int handleIdx)
  *Saves a backup of the current internal config (e.g. all the options specifiable by ConfigSet* routines), which can later be reloaded using digHoloConfigBackupLoad()*

- int digHoloConfigBackupLoad (int handleIdx)
  *Loads a backup of a previously saved internal config (e.g. all the options specifiable by ConfigSet* routines). Previous call to digHoloConfigBackupSave() is required.*

- void digHoloDebugRoutine (int handleIdx)
  *Not for general usage. A routine that serves as a place to quickly write debug code for testing purposes.*

- float * digHoloFrameSimulatorCreate (int frameCount, int *frameWidth, int *frameHeight, float *pixelSize, int *polCount, float **refTiltX, float **refTiltY, float **refDefocus, float **refWaist, float **refBeamCentreX, float **refBeamCentreY, complex64 **refAmplitude, int *beamGroupCount, float **beamWaist, complex64 **beamCoefs, float **beamCentreX, float **beamCentreY, int *cameraPixelLevelCount, int fillFactorCorrection, float **wavelengths, int *wavelengthCount, int wavelengthOrdering, int printToConsole, unsigned short **pixelBuffer16, const char *fname)
  *Routine for generating simulated digital holography frame data for testing and benchmarking purposes.*

- float * digHoloFrameSimulatorCreateSimple (int frameCount, int frameWidth, int frameHeight, float pixelSize, int polCount, float wavelength, int printToConsole)
  *A simplified version of the digHoloFrameSimulatorCreate() routine. Whereby the user specifies the minimum necessary parameters to generate frames, and everything else is default.*

- int digHoloFrameSimulatorDestroy (float *pixelBuffer)
  *Destroys the memory allocations associated with the digHoloFrameSimulatorCreate() call. A thin wrapper around a memory deallocation (free) call.*

## Detailed Description

Configuring how console printing is performed, plotting tools, debug tools, as well as additional utilities.

## Function Documentation

### int digHoloConfigBackupLoad (int *handleIdx*)

Loads a backup of a previously saved internal config (e.g. all the options specifiable by ConfigSet* routines). Previous call to digHoloConfigBackupSave() is required.

This is supposed to be a convienient utility to save a copy of the internal state of the digHoloObject, which can later be recalled if required.

**Parameters**

| in | *handleIdx* | : enumerated handle index of the digHoloObject |
|---|---|---|

**Returns**

errorCode : [DIGHOLO_ERROR_SUCCESS, DIGHOLO_ERROR_INVALIDHANDLE, DIGHOLO_ERROR_NULLPOINTER]. Null pointer would mean there is no backup config for the specified digHoloObject. Likely because digHoloConfigBackupSave() has never been called.

### int digHoloConfigBackupSave (int *handleIdx*)

Saves a backup of the current internal config (e.g. all the options specifiable by ConfigSet* routines), which can later be reloaded using digHoloConfigBackupLoad()

This is supposed to be a convienient utility to save a copy of the internal state of the digHoloObject, which can later be recalled if required.

**Parameters**

| in | *handleIdx* | : enumerated handle index of the digHoloObject |
|---|---|---|

**Returns**

errorCode : [DIGHOLO_ERROR_SUCCESS, DIGHOLO_ERROR_INVALIDHANDLE, DIGHOLO_ERROR_NULLPOINTER]. Null pointer would mean there is no current config for the specified digHoloObject.

### int digHoloConfigGetVerbosity (int *handleIdx*)

Returns how much information is printed to console/file during operation.

0 : Nothing printed except deep errors from underlying libraries (e.g. MKL). 1 : Basic info like errors, start/stop summaries. 2 : Debug-like printouts with relatively high level of detail. 3 : You will be in our thoughts and prayers.

**Parameters**

| in | *handleIdx* | : enumerated handle index of the digHoloObject |
|---|---|---|

**Returns**

verbosity : verbosity level. Invalid handle will return 0.

### int digHoloConfigSetVerbosity (int *handleIdx*, int *verbosity*)

Specifies how much information is printed to console/file during operation.

0 : Nothing printed except deep errors from underlying libraries (e.g. MKL). 1 : Basic info like errors, start/stop summaries. 2 : Debug-like printouts with relatively high level of detail. 3 : May God have mercy on your soul.

**Parameters**

| in | *handleIdx* | : enumerated handle index of the digHoloObject |
|---|---|---|
| in | *verbosity* | : verbosity level |

**Returns**

errorCode [DIGHOLO_ERROR_SUCCESS, DIGHOLO_ERROR_INVALIDHANDLE] Setting non-existent verbosity levels will not raise errors. Specifying less than 0 will default to 0. Specifying above the maximum level will behave like the maximum level.

### int digHoloConsoleRedirectToFile (char * *filename*)

Calling this function will redirect stdout console to the specified filename.

**Parameters**

| in | *filename* | : filename of the desired output text file. |
|---|---|---|

**Returns**

erroCode : [DIGHOLO_ERROR_SUCCESS, DIGHOLO_ERROR_COULDNOTCREATEFILE, DIGHOLO_ERROR_NULLPOINTER]

### int digHoloConsoleRestore ()

Sets the console to stdout.

**Returns**
erroCode : [DIGHOLO_ERROR_SUCCESS]

**void digHoloDebugRoutine (int  *handleIdx*)**

Not for general usage. A routine that serves as a place to quickly write debug code for testing purposes.

**Parameters**

| in | *handleIdx* | : enumerated handle index of the digHoloObject |
|---|---|---|

**float * digHoloFrameSimulatorCreate (int  *frameCount*, int *  *frameWidth*, int * *frameHeight*, float *  *pixelSize*, int *  *polCount*, float **  *refTiltX*, float **  *refTiltY*, float **  *refDefocus*, float **  *refWaist*, float **  *refBeamCentreX*, float **  *refBeamCentreY*, complex64 **  *refAmplitude*, int *  *beamGroupCount*, float **  *beamWaist*, complex64 **  *beamCoefs*, float **  *beamCentreX*, float **  *beamCentreY*, int *  *cameraPixelLevelCount*, int   *fillFactorCorrection*, float **  *wavelengths*, int *  *wavelengthCount*, int   *wavelengthOrdering*, int   *printToConsole*, unsigned short **  *pixelBuffer16*, const char *  *fname*)**

Routine for generating simulated digital holography frame data for testing and benchmarking purposes.

Null pointers can be passed into many of these parameters, in which case, the routine will initialise the parameters to some default value, and return a pointer to an array containing those default values.

Many of the input pointers are passed in by reference (pointer to the array pointer). Hence if the user specifies a pointer to a null pointer, after the routine is called, the pointer will now point to memory containing the default values selected by the routine.

**Parameters**

| in | *frameCount* | : The number of frames to generate |
|---|---|---|
| in,out | *frameWidth* | : The width of each frame in pixels. The value will potentially be modified to conform with being a multiple of DIGHOLO_PIXEL_QUANTA |
| in,out | *frameHeight* | : The height of each frame in pixels. The value will potentially be modified to conform with being a multiple of DIGHOLO_PIXEL_QUANTA |
| in,out | *pixelSize* | : The physical dimension of each pixel. If specified <0, this will return a default value of pixelSize |
| in,out | *polCount* | : The number of polarisation components. Will enforce conformance with >0 and <=DIGHOLO_POLCOUNTMAX |
| in,out | *refTiltX* | : Pointer to an array containing the desired tilts for each polarisation component. Passing a null pointer will result in default values. |
| in,out | *refTiltY* | : Pointer to an array containing the desired tilts for each polarisation component. Passing a null pointer will result in default values. |
| in,out | *refDefocus* | : Pointer to an array containing the desired defocus for each polarisation component. Passing a null pointer will result in default values. |
| in,out | *refWaist* | : Pointer to an array containing the desired Gaussian waist radius for the Reference wave in each polarisation component. Passing a null pointer will result in default values (plane-wave). |
| in,out | *refBeamCentreX* | : Pointer to an array containing the position of the Reference wave beam centre. Passing a null pointer will result in default values. |
| in,out | *refBeamCentreY* | : Pointer to an array containing the position of the Reference wave beam centre. Passing a null pointer will result in default values. |

| in,out | *refAmplitude* | : Pointer to an array containing a complex amplitude factor to be applied to each polarisation component of the Reference wave. |
|---|---|---|
| in,out | *beamGroupCount* | : The number of Hermite-Gaussian groups for the test basis. If set to <=0, the groupCount will be selected so as to create enough orthogonal modes to satisfy frameCount. |
| in,out | *beamWaist* | : Pointer to an array containing the beam radius for each polarisation component of the Hermite-Gaussian basis. Passing a null pointer will result in default values which approximately fill the frame. |
| in,out | *beamCoefs* | : Pointer to a frameCount x polCount x modeCount array of Hermite-Gaussian coefficients used to generate the fields for each frame. Passing a null pointer will result in default values. |
| in,out | *beamCentreX* | : Pointer to an array containing the beam centre for each polarisation component. Passing a null pointer will result in default values. |
| in,out | *beamCentreY* | : Pointer to an array containing the beam centre for each polarisation component. Passing a null pointer will result in default values. |
| in,out | *cameraPixelLevelCount* | : The total number of discrete intensity levels for the simulated frames. e.g. 14-bit camera simulation would be 2^14=16384 |
| in,out | *fillFactorCorrection* | : Defines whether or not a sinc envelope should be applied in Fourier space to the frames, simulating the effect of finite (100%) pixel fill factor. |
| in,out | *wavelengths* | : Pointer to an array of 'wavelengthCount' wavelengths. Specifying the wavelengths for each frame, as per the wavelengthOrdering parameter. Passing a null pointer will result in default values. |
| in,out | *wavelengthCount* | : The length of the wavelengths array. Specifying value<1 will result in default behaviour. |
| in,out | *wavelengthOrdering* | : Specifies the relationship between frames within the frameCount, and their corresponding wavelength in the wavelengths array. As per DIGHOLO_WAVELENGTHORDER_INPUT = DIGHOLO_WAVELENGTHORDER_FAST/SLOW |
| in | *printToConsole* | : Whether or not to print information to the console. |
| out | *pixelBuffer16* | : Pointer to an output array containing the frame data in uint16 format. |
| in | *fname* | : Path to a file to write the uint16 frame data as a binary file. Passing a null will result in no file being written. |

**Returns**

pixelBuffer : Pointer to an output array containing the frame data in float32 format.

**float * digHoloFrameSimulatorCreateSimple (int  *frameCount*, int  *frameWidth*, int  *frameHeight*, float  *pixelSize*, int  *polCount*, float  *wavelength*, int  *printToConsole*)**

A simplified version of the digHoloFrameSimulatorCreate() routine. Whereby the user specifies the minimum necessary parameters to generate frames, and everything else is default.

This function is also less forgiving than the digHoloFrameSimulatorCreate() routine. e.g. invalid parameters will cause the routine to halt and return a null pointer, rather than choosing default values. This function also takes values directly as arguments, rather than pointers.

**Parameters**

| in | *frameCount* | : The number of frames to generate |
|---|---|---|
| in | *frameWidth* | : The width of each frame in pixels. |
| in | *frameHeight* | : The height of each frame in pixels |
| in | *pixelSize* | : The physical dimension of each pixel. |
| in | *polCount* | : The number of polarisation components. |
| in | *wavelength* | : The operating wavelength. |

| in | *printToConsole* | : Whether or not to print information to the console. |
|---|---|---|

**Returns**

pixelBuffer : Pointer to an output array containing the frame data in float32 format. To destroy the memory created during this routine, call digHoloFrameSimulatorDestroy(pixelBuffer)

### int digHoloFrameSimulatorDestroy (float * *pixelBuffer*)

Destroys the memory allocations associated with the digHoloFrameSimulatorCreate() call. A thin wrapper around a memory deallocation (free) call.

Internally, the digHoloFrameSimulatorCreate*() routines allocate all persistent memory as a single chunk, which is then divied up and aliased internally. Hence freeing the float32 pixelBuffer, also frees the other pointers such as the uint16 pixelBuffer16 and the memory used to return parameters like tilt, beamwaist etc.

**Parameters**

| in | *pixelBuffer* | : Pointer to the array of pixels returned by the digHoloFrameSimulatorCreate() routine. |
|---|---|---|

**Returns**

errorCode : [DIGHOLO_ERROR_SUCCESS, DIGHOLO_ERROR_MEMORYALLOCATION, DIGHOLO_ERROR_NULLPOINTER]

### unsigned char * digHoloGetViewport (int *handleIdx*, int *displayMode*, int *forceProcessing*, int * *width*, int * *height*, char ** *windowString*)

Visualises internal buffers. Renders an RGB bitmap based on what type of plot you want to see (displayMode).

**Parameters**

| in | *handleIdx* | : enumerated handle index of the digHoloObject |
|---|---|---|
| in | *displayMode* | : type of plotting function to perform |
| in | *forceProcessing* | : Plot existing batch data (0), or run a new batch of processing before plotting (1) |
| out | *width* | : int32 pointer into which the width of the resulting plot will be stored |
| out | *height* | : int32 pointer into which the height of the resulting plot will be stored |
| out | *windowString* | : char* pointer to an internal buffer containing a string which can serve as a window title containing basic information. |

**Returns**

bitmap : A pointer to an internal RGB bitmap of the resulting plot.

### int digHoloGetViewportToFile (int *handleIdx*, int *displayMode*, int *forceProcessing*, int * *width*, int * *height*, char ** *windowString*, const char * *filename*)

Outputs the RGB pixel buffer returned by digHoloGetViewport to a binary file.

**Parameters**

| in | *handleIdx* | : enumerated handle index of the digHoloObject |
|---|---|---|
| in | *displayMode* | : type of plotting function to perform |
| in | *forceProcessing* | : Plot existing batch data (0), or run a new batch of processing before plotting (1) |

| out | *width* | : int32 pointer into which the width of the resulting plot will be stored |
|---|---|---|
| out | *height* | : int32 pointer into which the height of the resulting plot will be stored |
| out | *windowString* | : char* pointer to an internal buffer containing a string which can serve as a window title containing basic information. |
| in | *filename* | : filename to write the pixelBuffer to. By default, this is a binary file containing RGB pixel data. However if the filename ends in '.bmp', then the viewport will be written to a bitmap file. |

**Returns**

errorCode : [DIGHOLO_ERROR_SUCCESS DIGHOLO_ERROR_FILENOTCREATED DIGHOLOERROR_INVALIDHANDLE]

**int digHoloRunBatchFromConfigFile (char \*  *filename*)**

Loads settings from a tab-delimited text file and performs a batch processing run.

Intended mostly for using the digHolo library as an executable called from the command-line, and for debugging purposes.

**Parameters**

| in | *filename* | : string containing the text file path. |
|---|---|---|

**Returns**

errorCode : [DIGHOLO_ERROR_SUCCESS, DIGHOLO_ERROR_FILENOTFOUND, DIGHOLO_ERROR_NULLPOINTER, DIGHOLO_ERROR_ERROR]

# Viewport display modes

Types of plotting functions the GetViewport function supports. Mostly used for visually internal buffers such as the FourierPlane, reconstructed fields, etc.

## Macros

- #define [DIGHOLO_VIEWPORT_NONE](#) 0
- #define [DIGHOLO_VIEWPORT_CAMERAPLANE](#) 1
- #define [DIGHOLO_VIEWPORT_FOURIERPLANE](#) 2
- #define [DIGHOLO_VIEWPORT_FOURIERPLANEDB](#) 3
- #define [DIGHOLO_VIEWPORT_FOURIERWINDOW](#) 4
- #define [DIGHOLO_VIEWPORT_FOURIERWINDOWABS](#) 5
- #define [DIGHOLO_VIEWPORT_FIELDPLANE](#) 6
- #define [DIGHOLO_VIEWPORT_FIELDPLANEABS](#) 7
- #define [DIGHOLO_VIEWPORT_FIELDPLANEMODE](#) 8
- #define [DIGHOLO_VIEWPORT_COUNT](#) 9
- #define [DIGHOLO_VIEWPORT_NAMELENGTH](#) 1024

## Detailed Description

Types of plotting functions the GetViewport function supports. Mostly used for visually internal buffers such as the FourierPlane, reconstructed fields, etc.

The plotting functions can behave differently, depending on whether you are plotting an individual frame (batchCount=1) or a batch. In general, when plotting a batch consisting of multiple elements, the GetViewport function will typically plot the sum/average absolute value or intensity over the whole batch.

## Macro Definition Documentation

### #define DIGHOLO_VIEWPORT_CAMERAPLANE   1

Viewport plot camera plane.

### #define DIGHOLO_VIEWPORT_COUNT   9

The total number of supporting Viewport plotting modes.

### #define DIGHOLO_VIEWPORT_FIELDPLANE   6

Viewport plot the reconstructed field.

### #define DIGHOLO_VIEWPORT_FIELDPLANEABS   7

Viewport plot the absolute value of the reconstructed field.

### #define DIGHOLO_VIEWPORT_FIELDPLANEMODE   8

Viewport plot the reconstructed field as a superposition of supported HG basis modes.

### #define DIGHOLO_VIEWPORT_FOURIERPLANE   2

Viewport plot absolute value of the Fourier plane.

**#define DIGHOLO_VIEWPORT_FOURIERPLANEDB   3**

    Viewport plot absolute value on log scale of the Fourier plane.

**#define DIGHOLO_VIEWPORT_FOURIERWINDOW   4**

    Viewport plot the Fourier window.

**#define DIGHOLO_VIEWPORT_FOURIERWINDOWABS   5**

    Viewport plot the absolute value of the Fourier window.

**#define DIGHOLO_VIEWPORT_NAMELENGTH   1024**

    Not a display mode, but the maximum number of characters in a viewport window title.

**#define DIGHOLO_VIEWPORT_NONE   0**

    Viewport plot disabled mode.

# Console printout verbosity levels

Specifies the desired level of detail which should be printed to the console.

## Macros

- #define DIGHOLO_VERBOSITY_OFF   0
- #define DIGHOLO_VERBOSITY_BASIC   1
- #define DIGHOLO_VERBOSITY_DEBUG   2
- #define DIGHOLO_VERBOSITY_COOKED   3

## Detailed Description

Specifies the desired level of detail which should be printed to the console.

## Macro Definition Documentation

### #define DIGHOLO_VERBOSITY_BASIC   1

Basic info like errors, start/stop summaries.

### #define DIGHOLO_VERBOSITY_COOKED   3

May God have mercy on your soul.

### #define DIGHOLO_VERBOSITY_DEBUG   2

Debug-like printouts with relatively high level of detail.

### #define DIGHOLO_VERBOSITY_OFF   0

Nothing printed except deep errors from underlying libraries (e.g. MKL).

# Deprecated

Deprecated functions that should no longer be used, but are kept for compatibility purposes.

## Functions

- float ** digHoloConfigGetZernCoefs (int handleIdx)
  *Returns a pointer to an internal array containing Zernike coefficients for aberration correction.*

- int digHoloConfigGetZernCount (int handleIdx)
  *Returns the maximum number of Zernike coefficients in the array. i.e. one of the dimensions of ZernCoefs.*

## Detailed Description

Deprecated functions that should no longer be used, but are kept for compatibility purposes.

## Function Documentation

### float ** digHoloConfigGetZernCoefs (int *handleIdx*)

Returns a pointer to an internal array containing Zernike coefficients for aberration correction.

Only tilt and defocus are implemented, and the preferred method of interfacing with these is the "digHoloConfigGetTilt" "digHoloConfigSetTilt" routines.

**Parameters**

| in | *handleIdx* | : enumerated handle index of the digHoloObject |
|----|-------------|------------------------------------------------|

**Returns**

zernPointer : Pointer to internal array [polIdx x zernikeIdx] containing the Zernike coefficients for each polarisation component.

### int digHoloConfigGetZernCount (int *handleIdx*)

Returns the maximum number of Zernike coefficients in the array. i.e. one of the dimensions of ZernCoefs.

Only tilt and defocus are implemented, and the preferred method of interfacing with these is the "digHoloConfigGetTilt" "digHoloConfigSetTilt" routines.

**Parameters**

| in | *handleIdx* | : enumerated handle index of the digHoloObject |
|----|-------------|------------------------------------------------|

**Returns**

zernCount : the number of zernike coefficients supports. This is typically 6, up to and including Zernike order 2. [piston, tiltx,tilty,defocus,astigX, astigY]

# Index

INDEX